\documentclass[nojss]{jss}
\usepackage{booktabs}
\usepackage{graphicx}
\usepackage{amsmath,amssymb}

\newtheorem{defi}{Definition}
\DeclareMathOperator{\argmin}{argmin}

\newcommand{\IR}{\mathbb{R}}
\newcommand{\IC}{\mathbb{C}}
\newcommand{\IZ}{\mathbb{Z}}

\author{Tobias Kley\\Ruhr-Universit\"at Bochum}
\title{Quantile-Based Spectral Analysis in an Object-Oriented Framework
and a Reference Implementation in \proglang{R}: The \pkg{quantspec} Package}

\Plainauthor{Tobias Kley} %% comma-separated
\Plaintitle{An Object-oriented Framework for Quantile-based Spectral Analysis and a Reference Implementation in R: The quantspec Package} %% without formatting
\Shorttitle{\pkg{quantspec}: Quantile-based Spectral Analysis in \proglang{R}} %% a short title (if necessary)

\Abstract{Quantile-based approaches to the spectral analysis of time series have
recently attracted a lot of attention. Despite a growing literature that
contains various estimation proposals, no systematic methods for computing
the new estimators are available to date. This paper contains two main
contributions. First, an extensible framework for quantile-based spectral
analysis of time series is developed and documented using object-oriented
models. A comprehensive, open source, reference implementation of this
framework, the \proglang{R} package \pkg{quantspec}, was recently contributed to
CRAN by the author of this paper. The second contribution of the present paper
is to provide a detailed tutorial, with worked examples, to this \proglang{R}
package.
A reader who is already familiar with quantile-based spectral analysis and whose
primary interest is not the design of the \pkg{quantspec} package, but how to
use it, can read the tutorial and worked examples (Sections~\ref{sec:3}
and~\ref{sec:4}) independently.}

\Plainkeywords{time series, spectral analysis, periodogram, quantile regression, copulas, ranks, R, quantspec, framework, object-oriented design}
\Address{
  Tobias Kley\\
  Department of Mathematics\\
  Institute of Statistics\\
  44780 Bochum, Germany\\
  E-mail: \email{tobias.kley@ruhr-uni-bochum.de}\\
  URL: \url{http://www.ruhr-uni-bochum.de/mathematik3/team/kley.html}
}
\begin{document}

\section{A short introduction to quantile-based spectral analysis}
\label{sec:1}

\subsection{Laplace and copula cumulants and their spectral representation}
\label{sec:Lcck}

Quantification of serial dependence in a stationary process
$(X_t)_{t \in \IZ}$ is traditionally based on its autocovariance and
autocorrelation functions, which measure linear dependencies among observations
at different times. Periodicities of a time series are then most
commonly analyzed by decomposing the autocovariance function, into a sum of
sines and cosines. This approach is referred to as (ordinary) spectral analysis of
time series and has been known for decades. As a statistical method, it
has been investigated many times and is well understood. In the analysis of
centered Gaussian time series this approach is particularly attractive, because
the autocovariance function completely characterizes the distribution of the
underlying process. If that process is not Gaussian, ordinary spectral
analysis suffers from typical weaknesses of $L_2$-methods: it is lacking
robustness against outliers and heavy tails, and is unable  to capture important
dynamic features such as changes in the conditional shape (skewness, kurtosis),
time-irreversibility, or dependence in the extremes. In addition to this, only time
series with an existing second moment can be analyzed at all. All of this was
previously realized by many researchers, and various extensions and
modifications of the $L^2$-periodogram have been proposed to remedy those drawbacks.

Approaches to robustifying the traditional spectral methods against
outliers and deviations from the distributional assumptions were taken, among
others, by
\cite{KleinerEtAl1979,KluppelbergMikosch1994,Mikosch1998,Katkovnik1998,
HillMcCloskey2013}. To account for more general dynamic features
alternative spectral concepts and tools were recently proposed.
A first step in that direction was taken by \citet{Hong1999}. In order to obtain
a complete description of the two-dimensional distributions at lag~$k$, he
introduced a  {\it generalized spectrum} where covariances ${\rm Cov}(X_t,
X_{t-k})$ are replaced by covariances ${\rm Cov} ({\rm e}^{{\rm i}x_1 X_t}, {\rm
e}^{{\rm i}x_2 X_{t-k}})$ yielding a spectrum closely related to the joint
characteristic functions of the pairs $(X_t, X_{t-k})$.
In the quantile-based approach to spectral analysis the objects of interest are
the \emph{Laplace cross-covariance kernel} \[\gamma_k(q_1, q_2) :=
\text{Cov}\Big(I\{ X_t \leq q_1\}, I\{ X_{t-k} \leq q_2\}\Big), \quad  q_1, q_2 \in \bar{\IR}, \ k \in \IZ,\] and the \emph{copula cross-covariance kernel} \[\gamma^{U}_k(\tau_1, \tau_2) := \Big(I\{ F(X_t) \leq \tau_1\}, I\{ F(X_{t-k}) \leq \tau_2\}\Big), \quad  \tau_1, \tau_2 \in [0,1], \ k \in \IZ,\]
where $I\{A\}$ denotes the indicator function of the event $\{A\}$ and $F$ the
marginal distribution function (the distribution function of any $X_t$, due to
the assumed stationarity).
Obviously these measures exist without the
necessity to make assumptions about moments. Also, when the underlying process
is not Gaussian, and the quantile-based measures of serial dependence are
considered as functions with arguments $q_1, q_2$, or $\tau_1, \tau_2$
respectively, they provide a much richer picture about the pairwise dependence
than would the autocovariances. As in the approach of \cite{Hong1999}, a
complete description of the joint distributions (or copulas) of the pairs $(X_t, X_{t-k})$
is available. A particular advantage of the copula cross-covariance kernel is
its invariance to monotone transformations. This allows to disentangle the
serial features from the marginal features.
For a full list of the properties and advantages of those dependence
measures the interested reader be refered to
\cite{Hong2000,Li2008,Li2012,Li2013,Li2014,Hagemann2011,LeeRao2012,DetteEtAl2013,KleyEtAl2014}
and \cite{Kley2014a}.

Under sumability conditions on $(\gamma_k)$ and $(\gamma_k^U)$ the
representations of $(\gamma_k)$ and $(\gamma_k^U)$ in the ``frequency domain''
take the form of the \emph{Laplace spectral density kernel}
\begin{equation}
  \label{def:LapSD}
  \mathfrak{f}_{q_1, q_2}(\omega) := \frac{1}{2\pi} \sum_{k=-\infty}^{\infty} \gamma_k(q_1, q_2) \text{e}^{-\text{i} k \omega}, \quad q_1, q_2 \in \bar{\IR}, \ \omega \in \IR,
\end{equation}
and the \emph{copula spectral density kernel}
\begin{equation}
  \label{def:CopSD}
  \mathfrak{f}_{q_{\tau_1}, q_{\tau_2}}(\omega) := \frac{1}{2\pi} \sum_{k=-\infty}^{\infty} \gamma_k^U(q_1, q_2) \text{e}^{-\text{i} k \omega}, \quad \tau_1, \tau_2 \in [0,1], \ \omega \in \IR,
\end{equation}
where $q_{\tau} := F^{-1}(\tau)$. By the relation
\[\gamma_k(q_1, q_2) = \int_{-\pi}^{\pi} \text{e}^{\text{i} k \omega} \mathfrak{f}_{q_1, q_2}(\omega) \text{d}{\omega},\]
and a similar representation for $\gamma_k^U(\tau_1, \tau_2)$, the
representations in the ``frequency domain'' are seen to be equivalent to the
``time domain'' quantities.

Sometimes considering the cumulated Laplace or copula spectral density kernels, which can be defined as
\begin{equation}
  \label{def:cumLapSD}
  \mathfrak{F}_{q_1, q_2}(\omega) := \int_0^{\omega} \mathfrak{f}_{q_1, q_2}(\lambda) {\rm d} \lambda, \quad
  q_1, q_2 \in \bar{\IR}, \ \omega \in [0,2\pi],
\end{equation}
and
\begin{equation}
  \label{def:cumCopSD}
  \mathfrak{F}_{q_{\tau_1}, q_{\tau_2}}(\omega) := \int_0^{\omega} \mathfrak{f}_{q_{\tau_1}, q_{\tau_2}}(\lambda) {\rm d} \lambda, \quad
  \tau_1, \tau_2 \in [0,1], \ \omega \in [0,2\pi],
\end{equation}
is more convenient.

Quantities as $\gamma_k$ and $\gamma^{U}_k$, and their spectral
representations~\eqref{def:LapSD}--\eqref{def:cumCopSD} naturally come into the
picture when the {\it clipped processes} $(I\{X_t \leq q\})_{t \in \IZ}$ and
$(I\{F(X_t) \leq \tau\})_{t \in \IZ}$ are investigated. Such binary processes
have been considered earlier in the literature by e.\,g.~\citet{Kedem1980}.
Observe that the quantile-based spectral quantities can be interpreted in terms
of an orthogonal increment process of a spectral representation of the clipped
process which exists for every stictly stationary process; no assumptions about
moments are necessary.

Recently, there has been a surge of interest in that type of concept,
with the introduction, under the names of \textit{Lap\-lace-, quantile-} and
\textit{copula} spectral density and spectral density {\it kernels}, of various
quantile-related spectral concepts, along with the corresponding sample-based
periodograms and smoothed periodograms
[cf.
\cite{Li2008,Li2012,Li2013,Li2014,Hagemann2011,LeeRao2012,DetteEtAl2013,Kley2014a,KleyEtAl2014}].

Despite the vast amount of theoretical work, a public software solution was so
far not available.

\subsection{Estimators for the quantile-based spectral analysis of time series}

In this section of the introduction, various estimators (the so-called quantile
periodograms) for the Laplace and copula spectra defined in
Section~\ref{sec:Lcck} are briefly considered.
For the upcoming definitions denote by $X_0, \ldots, X_{n-1}$ an observed time
series of the process $(X_t)_{t \in \IZ}$, by
\[\hat F_n(x) := \sum_{t=0}^{n-1} I\{X_t \leq x\}\]
the \emph{empirical distribution function} of $X_0, \ldots, X_{n-1}$, and by
$\Re z$ and $\Im z$ the real and imaginary part of $z = \Re z + \text{i} \Im z
\in \IC$, respectively.
Further,
\[\rho_\tau (x):= x(\tau - I\{  x \leq 0\}) = (1-\tau ) \vert x\vert I\{ x\leq
0\} + \tau   \vert x\vert I\{ x > 0\},\]
denotes the so-called \emph{check function} [cf.~\citet{Koenker2005}].

\begin{defi}[Quantile-regression based periodograms]
  \label{def:qregPG} ~\\
  For $\omega \in \IR$ and $\tau_1, \tau_2 \in (0,1)$ the
  \emph{Laplace periodogram} $\hat L_{n}^{\tau_1, \tau_2}(\omega)$, and the
  \emph{rank-based Laplace periodogram} $\hat L_{n,R}^{\tau_1,\tau_2}(\omega)$ are defined as
  \begin{equation*}
    \hat L_{n}^{\tau_1, \tau_2}(\omega) := (2\pi n)^{-1} \hat
    b_{n}^{\tau_1}(\omega) \hat b_{n}^{\tau_2}(-\omega), \quad \hat
    L_{n,R}^{\tau_1,\tau_2}(\omega) := (2\pi n)^{-1} \hat b_{n,R}^{\tau_1}(\omega) \hat b_{n,R}^{\tau_2}(-\omega),
  \end{equation*}
  where, for $\omega \neq 0 \mod \pi$, and $\tau\in (0,1)$,
  \begin{align}
  (\hat a_{n}^{\tau}(\omega), \hat b_{n}^{\tau}(\omega)) & := \argmin_{a \in \IR, b \in \IC} \sum_{t=0}^{n-1} \rho_{\tau}(n X_{t} - a - 2 \cos(\omega t) \Re b + 2 \sin(\omega t) \Im b), \label{L1regre} \\
  (\hat a_{n,R}^{\tau}(\omega), \hat b_{n,R}^{\tau}(\omega)) & := \argmin_{a \in \IR, b \in \IC} \sum_{t=0}^{n-1} \rho_{\tau}(n \hat F_n(X_t) - a - 2 \cos(\omega t) \Re b + 2 \sin(\omega t) \Im b).\nonumber
  \end{align}
  If $\omega = 0 \mod \pi$, when the regressor that yields the imaginary part of the estimate vanishes, the estimates need to be adapted as follows: for $\omega_{\pi} = 2\pi (j+1/2)$, $j \in \IZ$, let
  \begin{align}
  (\hat a_{n}^{\tau}(\omega_{\pi}), \hat b_{n}^{\tau}(\omega_{\pi})) & := \argmin_{a \in \IR, b \in \IR} \sum_{t=0}^{n-1} \rho_{\tau}(n X_t - a - \cos(\omega_{\pi} t) b), \nonumber \\
  (\hat a_{n,R}^{\tau}(\omega_{\pi}), \hat b_{n,R}^{\tau}(\omega_{\pi})) & := \argmin_{a \in \IR, b \in \IR} \sum_{t=0}^{n-1} \rho_{\tau}(n \hat F_n(X_t) - a - \cos(\omega_{\pi} t) b).\nonumber
  \end{align}
\end{defi}

For $\omega \in 2\pi \IZ$ an adaptation is also possible [cf.~\cite{Kley2014a}],
but since it is not required for the definition of the smoothed estimates it is
omitted for the sake of brevity. Observe that the rank-based periodograms
obtained their name due to the fact that $n \hat F_n(X_t)$ is the rank of $X_t$
among $X_0, \ldots, X_{n-1}$

The Laplace periodograms trace back to \cite{Katkovnik1998} who, in the field of
signal processing, suggested $L_p$ estimators in a harmonic linear model.
\cite{Li2008} proved asymptotic normality of the Laplace periodograms for
$\tau_1 = \tau_2 = 0.5$ and later extended the approach to arbitrary
quantiles with $0 < \tau_1 = \tau_2 < 1$ \citep{Li2012}.
\cite{DetteEtAl2013} introduced the estimator with distinct quantile levels
$\tau_1$ and $\tau_2$ (not necessarily equal), and also considered the
rank-based version.

Another estimator is based on the discrete Fourier transform of clipped
processes and can be defined as follows:
\begin{defi}[Periodograms based on clipped time series]
  \label{def:clippedPG} ~\\
  For $\omega \in \IR$ and $q_1, q_2 \in \IR$, the
  \emph{clipped time-series periodogram} is defined as
  \begin{equation*}
    I_{n}^{q_1, q_2}(\omega) := (2\pi n)^{-1} d_{n}^{q_1}(\omega) d_{n}^{q_2}(-\omega), \quad
    d_{n}^{q}(\omega) := \sum_{t=0}^{n-1} I\{X_t \leq q\} {\rm e}^{-{\rm i} \omega t}.
  \end{equation*}
  For $\omega \in \IR$ and $\tau_1, \tau_2 \in [0,1]$ the \emph{copula rank
  periodogram} (for short CR periodogram) is defined as
  \begin{equation*}
    \label{eq:inr}
    I_{n,R}^{\tau_1,\tau_2}(\omega) := (2\pi n)^{-1} d_{n,R}^{\tau_1}(\omega) d_{n,R}^{\tau_2}(-\omega), \quad
    d_{n,R}^{\tau}(\omega) := \sum_{t=0}^{n-1} I\{\hat F_n(X_t) \leq \tau\}
    {\rm e}^{-{\rm i} \omega t}.
  \end{equation*}
\end{defi}

Note the similarity between all the quantile periodograms and the
cross-perio\-do\-grams in multivariate time series analysis [cf., e.\,g., \cite{Brillinger1975}, p.\,235]: each periodogram is a product of two frequency representation objects computed at frequencies ($\omega$ and $-\omega$) that sum to zero.

The estimator based on the discrete
Fourier transformation of clipped time series was introduced by \cite{Hong2000},
who used it for a test of pairwise independence. \cite{Hagemann2011} analyzed
the special case of $\tau_1 = \tau_2$ in the presence of serial dependence.
The case of distinct quantile levels $\tau_1$ and $\tau_2$ (not necessarily
equal) was discussed in \cite{KleyEtAl2014}, where weak convergence to a
Gaussian process was established.
\citet{LeeRao2012} investigated the distributions of Cram\' er-von Mises type
statistics, based on empirical joint distributions.

As in the traditional case the new periodograms are not consistent estimators
[cf. the positive variances of the limit distributions in Theorems~3.2 and~3.4
in \cite{DetteEtAl2013} or Proposition~3.4 in \cite{KleyEtAl2014}].

\subsection{Smoothing the quantile periodograms}
\label{sec:SmEstrs}

To achieve
consistency of the estimators we convolve the sequence of periodograms (indexed
with the Fourier frequencies) with a sequence of weighting functions $W_n$.
The smoothed periodograms are defined as follows:

\begin{defi}[Smoothed quantile-regression based
periodograms]\label{def:SmqregPG} ~\\
  For $\omega \in \IR$ and $\tau_1, \tau_2 \in (0,1)$ the \emph{smoothed Laplace
  periodogram} $\hat{\mathfrak{f}}_{n}(\tau_1, \tau_2; \omega)$ and
  \emph{smoothed rank-based Laplace periodogram}
  $\hat{\mathfrak{f}}_{n,R}(\tau_1, \tau_2; \omega)$ are defined as
  \begin{equation*}
  \begin{split}
    \hat{\mathfrak{f}}_{n}(\tau_1, \tau_2; \omega) & := \frac{2\pi}{n} \sum_{s=1}^{n-1} W_n\big( \omega - 2\pi s / n \big) \hat L_{n}^{\tau_1, \tau_2}(2 \pi s / n), \\
    \hat{\mathfrak{f}}_{n,R}(\tau_1, \tau_2; \omega) & := \frac{2\pi}{n} \sum_{s=1}^{n-1} W_n\big( \omega - 2\pi s / n \big) \hat L_{n,R}^{\tau_1, \tau_2}(2 \pi s / n).
  \end{split}
  \end{equation*}
\end{defi}

\begin{defi}[Smoothed periodograms based on clipped time
series]\label{def:SmclippedPG} ~\\
  For $\omega \in \IR$ and $q_1, q_2 \in \IR$ the \emph{smoothed clipped time
  series periodogram} is defined as
  \begin{equation*}
    \hat G_{n}(q_1, q_2; \omega) := \frac{2\pi}{n} \sum_{s=1}^{n-1} W_n\big( \omega - 2\pi s / n \big) I_{n}^{q_1, q_2}(2 \pi s / n).
  \end{equation*}
  For $\omega \in \IR$ and $\tau_1, \tau_2 \in [0,1]$ the \emph{smoothed copula
  rank periodogram} is defined as
  \begin{equation*}
    \hat G_{n,R}(\tau_1, \tau_2; \omega) := \frac{2\pi}{n} \sum_{s=1}^{n-1} W_n\big( \omega - 2\pi s / n \big) I_{n,R}^{\tau_1, \tau_2}(2 \pi s / n).
  \end{equation*}
\end{defi}

When
the weight functions are such that with $n \rightarrow \infty$ only the weights
in a shrinking neighborhood of zero will be positive, the estimators will be
consistent [cf. Theorem~3.7 in \cite{Kley2014a}]. Under suitable assumptions,
scaled versions of $\hat G_n(\cdot, \cdot; \omega)$ and $\hat
G_{n,R}(\cdot, \cdot; \omega)$ converge weakly to complex-valued
Gaussian processes [cf. Theorem 3.5 and 3.6 in \cite{KleyEtAl2014}]. A comprehensive description of all estimators and their
asymptotic properties is available in \cite{Kley2014a}.

\section{Conceptual design of the framework}
\label{sec:2}

\subsection{An analysis of functional requirements}

The \pkg{quantspec} software project was triggered by the development of the quantile-based methods for spectral analysis [cf.~Section~\ref{sec:1}, \cite{DetteEtAl2013} and \cite{KleyEtAl2014}]. The primary aim has been to make these new methods accessible to a wide range of users.

Before going into the programming-specific details of the project, a conceptual design, non-specific to any specific programming language was developed. By this procedure, additional insight and a thorough documentation of the computational characteristics of quantile-based spectral analysis could be gained. The conceptual design serves as a blueprint for implementations in (possibly) various environments and can easily be transformed into an implementation plan including the details specific to the respective programming environment.

Aiming for a software system that is most flexible in the ways in which it can be used, that can easily be extended in functionality and also for the ease of its maintenance an object-oriented design was chosen. This type of design also contributes to a structure of the system that can be better understood, both by users and developers. The general structure of the system for performing quantile-based spectral analysis is described using class diagrams of the unified modeling language (UML). In these diagrams, all necessary components and their interrelations are described in a formal manner.

To understand the specification of the framework, recall that in an
object-oriented design the components of the system are objects encapsulating
both data and behavior of a specific ``real-world'' entity. The structure of
each object can thus be described by a meaningful name (the \emph{class name}),
a collection of data fields (in \proglang{R} these are called \emph{slots}) and
implementations of the behavior (in \proglang{R} these implementations are
called \emph{methods}). In a class diagram each class (i.\,e., the composite of
class name, data files and implemented behavior) is represented as a rectangle
subdivided into three blocks. The name of the class is given in the top block,
the data fields in the middle block and the methods in the bottom block. Note
that in the unified modeling language the data fields and methods are specified
in a standardized format. In this format the first symbol is an abbreviation
used to specify the visibility of the class member. Here ``+'' for public and
``--'' for private members are used, meaning that the member is intended to be
seen (and used) from outside the object or from inside the object only,
respectively. For a data field the name is then followed by a colon and the type
of the field. For a method the parameters are given in parenthesis; optional
parameters are denoted by two dots. In the class diagram, relationships between
classes are marked as lines connecting them. Currently two different types of
relationships are modeled. A line with a triangular shaped tip at one end is
used to declare a \emph{generalization} relationship (sometimes also coined
inheritance or ``is a'' relationship). The class at the end of the line with the
triangle is called the \emph{superclass} or the parent, while the class on the other side is called the \emph{subclass} or the child. In particular this type of relationship implies that an object that is an instance to the subclass and therefore provides all the subclasses' data fields and methods will also provide the data fields and methods of the superclass (and the superclasses' superclass it there are such, etc.). The second type of relationships used in this framework is that of an \emph{aggregation} (sometimes called ``has a'' relationship). A line with a hollow diamond at one end is used to denote this kind of relationship, where objects to the class at the end of the line with the diamond are the ones having objects of the class at the other end of the line as a part of them. At each end of the line the so-called cardinalities are denoted in the form of two numbers with dots in between them. The left number is the minimum number of objects of that type in the relationship that need to exist, the right number is the maximum number. A star is used to denote an unknown positive integer. If min and max cardinality coincide they are displayed as one number without the dots.

For the class diagrams in this manuscript the classes are arranged in a way such that (whenever possible) generalization relationships are displayed with the superclass on top and the subclasses in the bottom. Aggregations are shown with aggregated classes to the left and/or the right of the class representing ``the whole''.

A graphical representation of the framework for quantile-based spectral analysis
is not given in one holistic diagram, but in two class diagrams that are on
display in Figures~\ref{fig:classdiagram1} and~\ref{fig:classdiagram2}. The structure of the framework is presented in two, thematically organized class diagrams rather than in one, because the 13 classes and their relations could not be fitted easily onto one page without breaking the above mentioned layout guidelines. On the other hand it was easy to group the classes by two topics.

In the next sections, all classes of the framework and their relations are going to be thoroughly described and motivated.

\subsection[The base class QSpecQuantity and its successors]{The base class \code{QSpecQuantity} and its successors}

Many of the quantities important for the quantile-based spectral analysis of a stationary time series [i.\,e., the estimators of Definitions~\ref{def:qregPG}--\ref{def:SmclippedPG} and the model quantities~\eqref{def:LapSD}--\eqref{def:cumCopSD}] are of the functional form,
\[Q_b: F \times T_1 \times T_2 \rightarrow \mathbb{C}, \quad b=1,\ldots,B,\]
where $F \subset \mathbb{R}$ is a set of frequencies [e.\,g, $F = [0,2\pi)$] and $T_1, T_2 \subset \bar{\mathbb{R}}$ are sets of levels. To provide a common interface to these objects the abstract class \code{QSpecQuantity} was introduced. Its data fields (i.\,e., the array \code{values}, a vector of reals \code{frequencies} and a list with two vectors of reals \code{levels}), are designed to store the sets
\begin{equation*}
\begin{split}
  \text{\code{frequencies}} & := \{\omega_1, \ldots, \omega_J\} \subset F, \\
  \text{\code{levels[[1]]}} & := \{q_{1,1}, \ldots, q_{1,K_1}\} \subset T_1, \\
  \text{\code{levels[[2]]}} & := \{q_{2,1}, \ldots, q_{2,K_2}\} \subset T_2, \\
\end{split}
\end{equation*}
and the family
\begin{equation*}
  \text{\code{values}} := \big( Q_b(\omega_j, q_{1,k_1}, q_{2,k_2})\big)_{j=1,\ldots,J; k_1=1,\ldots, K_1; k_2=1,\ldots, K_2; b=1,\ldots,B}.
\end{equation*}

Note that the handling of a family of $B$ quantile spectral quantities is necessary when bootstrapping replicates are present. The special case of only one function $Q$ can easily be handled by setting $B=1$.

There are four classes inheriting the data structure and the method \code{show}%
\footnote{The function show is used for printing objects of this class, and all superclasses, to the console.}
from the abstract class \code{QSpecQuantity}. Two such classes, \code{QuantilePG} and \code{SmoothedPG}, implement the computation of the various quantile periodograms and smoothed quantile periodograms, respectively.  A more detailed description has to include the other related classes and is therefore deferred to a separate section (i.\,e., Section~\ref{sec:2-2}). A graphical representation of the relevant parts of the framework can be seen in Figure~\ref{fig:classdiagram1}. The other two of the classes generalizing the abstract class \code{QSpecQuantitiy} are referred to by the names \code{QuantileSD} and \code{IntegrQuantileSD}. These two classes implement the model quantities~\eqref{def:LapSD}--\eqref{def:cumCopSD}. The graphical representation can be seen in Figure~\ref{fig:classdiagram2}. A detailed description is deferred to Section~\ref{sec:2-3}.

\subsection{Implementation of the quantile-based (smoothed) periodograms}
\label{sec:2-2}

\begin{figure}[t]
  \centering
  \scalebox{1.2}{\includegraphics{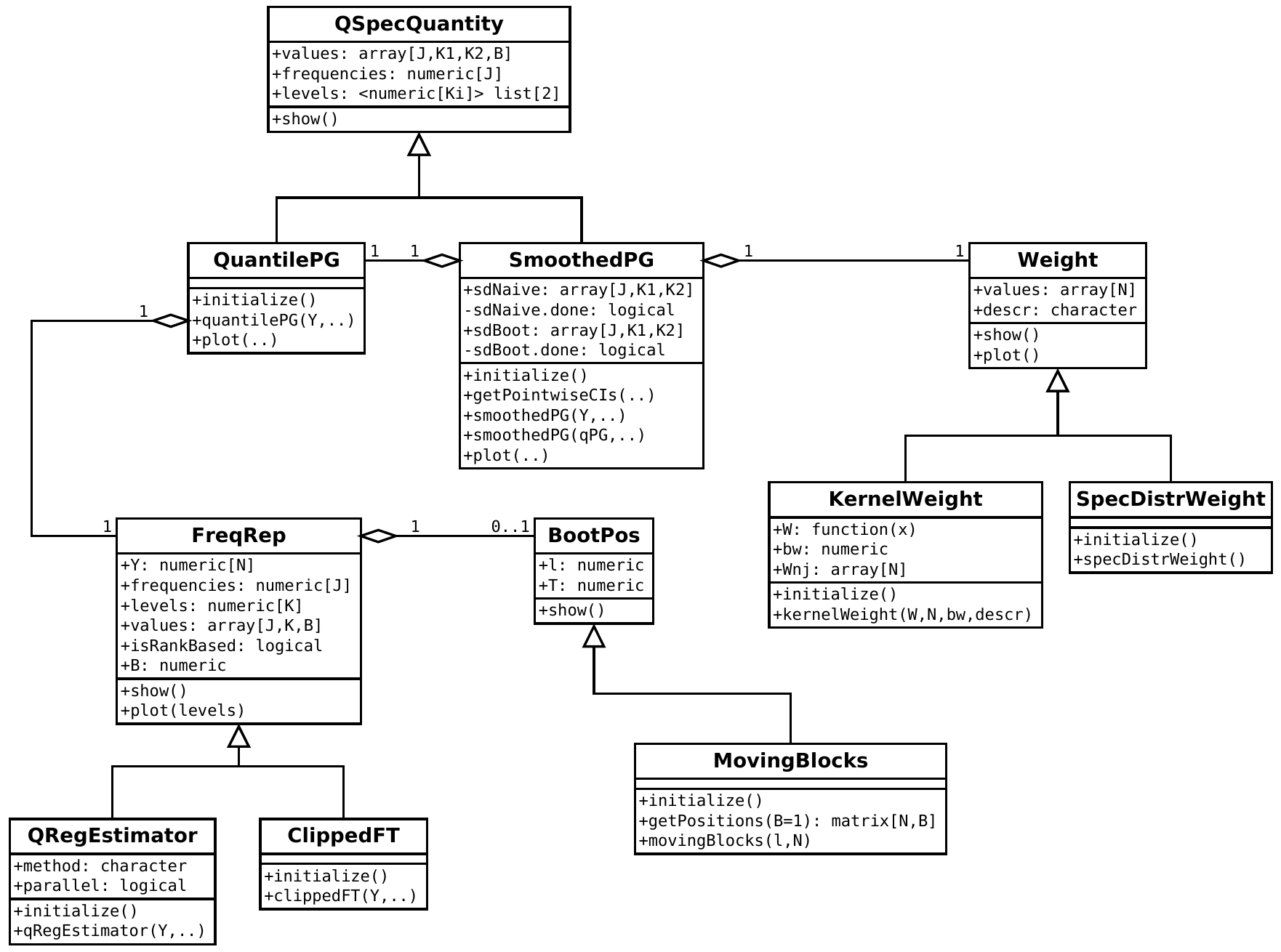}}
  \caption{Classes implementing the quantile-based periodograms and smoothed
  periodograms.}
  \label{fig:classdiagram1}
\end{figure}

The components relevant to the implementation of the quantile-based spectral statistics are presented in Figure~\ref{fig:classdiagram1}. As alluded to in the previous section the two classes \code{QuantilePG} and \code{SmoothedPG} will do the job, in conjunction with the superclass \code{QSpecQuantity} from which they inherit the data structure to store the computed values.

To better understand the implementation surrounding \code{QuantilePG}, observe that the quantile-based periodograms defined in Definitions~\ref{def:qregPG} and~\ref{def:clippedPG} share the common structure of an outer product (scaled with $(2\pi n)^{-1}$). To compute either one of the four periodograms
\begin{equation*}
\begin{split}
\hat L_{n,R}^{\tau_1,\tau_2}(\omega), \
\hat L_{n}^{\tau_1, \tau_2}(\omega), \
I_{n,R}^{\tau_1,\tau_2}(\omega), & \quad \tau_1 \in T_1, \ \tau_2 \in T_2, \ \omega \in F, \\
I_{n}^{q_1, q_2}(\omega) & \quad q_1 \in Q_1, \ q_2 \in Q_2, \ \omega \in F,
\end{split}
\end{equation*}
it suffices to perform the same operation to one of the frequency representation objects
\begin{equation}
\label{eqn:FR}
\begin{split}
\hat b_{n,R}^{\tau}(\omega), \
\hat b_{n}^{\tau}(\omega), \
d_{n,R}^{\tau}(\omega), & \quad \tau \in T_1 \cup T_2, \ \omega \in F, \\
d_{n}^{q}(\omega) & \quad q \in Q_1 \cup Q_2, \ \omega \in F,
\end{split}
\end{equation}
respectively. In the framework this fact is incorporated by introduction of the abstract class \code{FreqRep} and its two subclasses \code{ClippedFT} and \code{QRegEstimator}, where the actual computations are implemented via the method \code{initialization()}. The class \code{FreqRep} serves as a common interface to the quantities in~\eqref{eqn:FR}. It provides data fields to store various information, including
\begin{itemize}
  \item the observations \code{Y} from which the quantities were computed,
  \item the \code{frequencies} and \code{levels} for which the computation was performed,
  \item the result of the computation, which is stored in an array \code{values}.
\end{itemize}
Further more, a flag \code{isRankBased} indicates whether, prior to the main computations, the observations $(X_t)$ were transformed to pseudo-observations $(\hat F_n(X_t))$. Performing this extra step will yield $\hat b_{n,R}^{\tau}(\omega)$ instead of $\hat b_{n}^{\tau}(\omega)$ or $d_{n,R}^{\tau}(\omega)$ instead of $d_{n}^{\tau}(\omega)$, respectively.
The class \code{BootPos} allows for performing a block bootstrap procedure by ``shuffling'' the observations and repeatedly doing the computations on these bootstrapped observations. Currently only one method, the \code{MovingBlocks} bootstrap, is implemented.

Now, turning attention to the class \code{SmoothedPG}, recall that the various
smoothed periodograms are all defined similarly, in the sense that computing the smoothed periodogram for $\omega_j := 2\pi j /n$, $j=1, \ldots, n-1$ basically means to do a discrete convolution of the sequence of quantile periodograms computed at $\omega_s$ with a sequence of appropriately chosen weight functions $W_n(\omega_s)$. Hence, everything needed for the smoothed periodogram is these two ingredients, which is reflected in the framework by two aggregation relationships involving the class \code{SmoothedPG}. The first such relationship links \code{SmoothedPG} to the \code{QuantilePG}s to be smoothed. The second such relationship links \code{SmoothedPG} to a class \code{Weight}, which provides a common interface to different weight functions. Currently two implementations are included. Employing weights of type \code{KernelWeight}, defined by a kernel \code{W} and a scale parameter~\code{bw} (bandwidth), will yield an estimator for the quantile (i.\,e, Laplace or copula) spectral density. An alternative is to use weights of type \code{SpecDistrWeight}, which yields estimators for the integrated quantile (i.\,e, Laplace or copula) spectral density.

\subsection{Implementation of the quantile-based spectral measures}
\label{sec:2-3}

\begin{figure}[t]
  \centering
  \scalebox{1.2}{\includegraphics{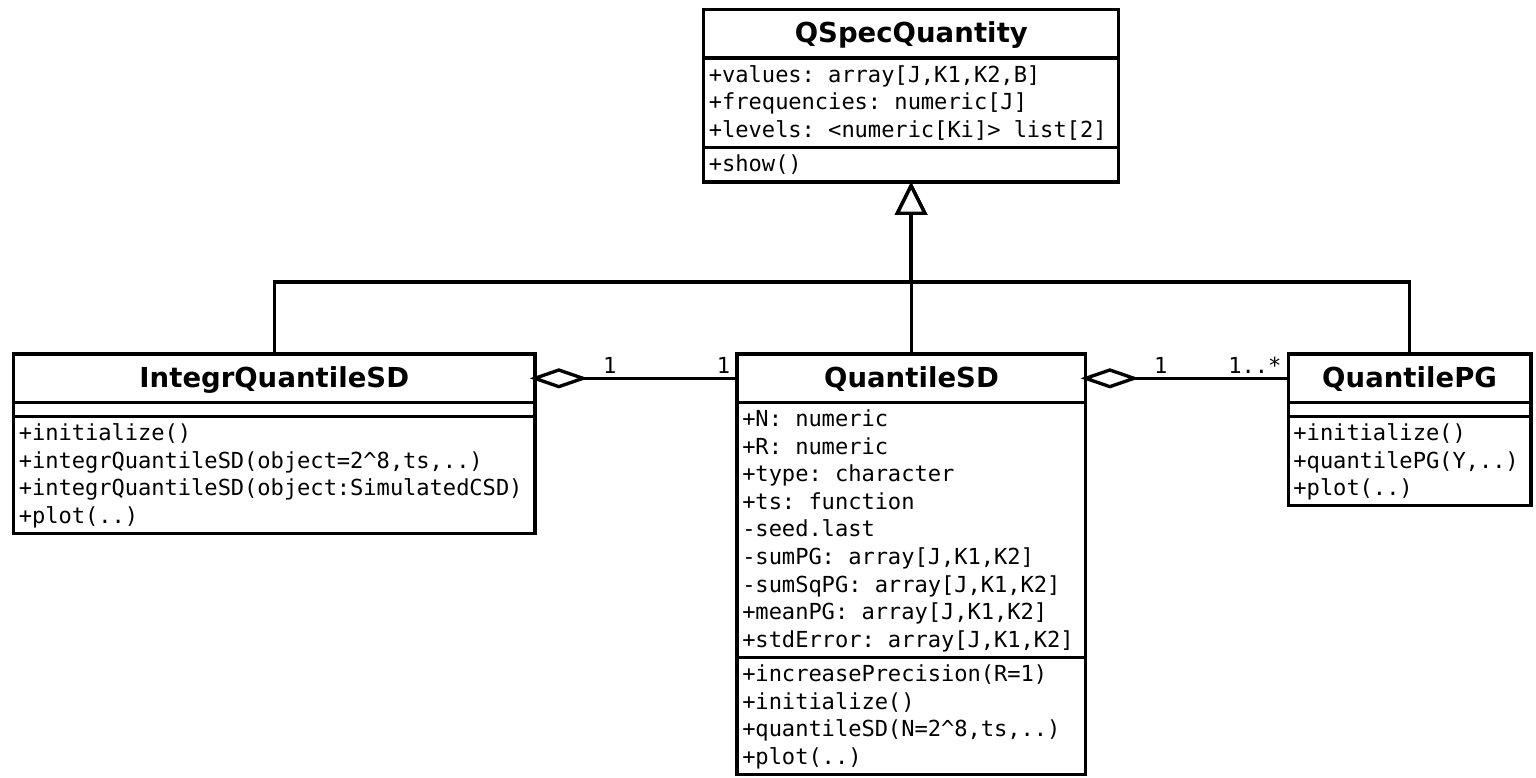}}
  \caption{Classes for the numerical computation of (integrated) copula and
  Laplace spectral densities via simulation.} % and their relations to each other
  \label{fig:classdiagram2}
\end{figure}

The classes~\code{QuantileSD} and \code{IntegrQuantileSD} were introduced to the framework to make the quantities~\eqref{def:LapSD}--\eqref{def:cumCopSD} available to the user.

To obtain access to a quantity of the form~\eqref{def:LapSD} or~\eqref{def:CopSD}, an instance of~\code{QuantileSD} can be created. In this case, a number of \code{R} independent copies of a time series of length \code{N} are obtained by calling the function \code{ts}. The function \code{ts} is a parameter to specify the model for which to obtain the model quantity and it handles the simulation process. Then, for each of these time series a~\code{QuantilePG} object is created and their values are averaged: first across the \code{R} independent copies, saving the result to \code{meanPG} and an estimation of the standard error to \code{stdError}. After that the averages are averaged again for each combination of levels across frequencies by smoothing. The result is then made available via the data field \code{values} of the superclass. By a call to the method \code{increasePrecision} the number of independent copies can be increased at any time to yield a less fluctuating average.

To obtain access to quantities of the form~\eqref{def:cumLapSD} or~\eqref{def:cumCopSD}, an instance of~\code{IntegrQuantileSD} can be created. For the computation an object of type~\code{QuantileSD} is created and subsequently the integral is approximated via a Riemann sum.

\section[Reference implementation: The R package quantspec]{Reference implementation: The \proglang{R} package \pkg{quantspec}}
\label{sec:3}
\subsection{Overview}

The \pkg{quantspec} package \citep{Kley2014} is intended to be used both by
theoretically oriented statisticians and also by data analysts, who work on a more applied
basis.
In order to address this broad group of potential users the \proglang{R} system for
statistical computing \citep{R2012} was chosen as a platform. The \proglang{R}
system is particularly well suited for the realization of this project, because
it is accessible from many operating systems, without charge, already available
to the targeted audience and, in particular, allows to integrate the package's
functionality among many other, well developed packages. An important example is
that the function \code{rq} of the \pkg{quantreg} package \citep{quantreg} could
be used.

Both \proglang{R} and the \pkg{quantspec} package are available from the
comprehensive \proglang{R} archive network (CRAN,
\url{http://cran.r-project.org}). The package's development is actively continued with the source code available from a GitHub repository (\url{https://github.com/tobiaskley/quantspec}). Besides the source code of the releases, which are also available from the CRAN servers, the GitHub repository additionally contains a detailed history of all changes, including comments, that were applied to the source code since April/10/2014, when the GitHub repository was created. The repository is organized into several branches: the \code{master} branch, the \code{develop} branch and possibly several topic branches. Since version 1.0-0 the \code{master} branch contains the source code of all release candidates and official releases, the \code{develop} branch contains the most recent updates and bug fixes that were not yet released. The topic branches contain the source code in which extensions to the package are developed.

To install a package from the source code straight from the repository use the
\code{install_github} function of the \pkg{devtools} package~\citep{devtools}. More precisely, install the \pkg{devtools} package, if it's not already installed, and call

\begin{Code}
R> devtools::install_github("tobiaskley/quantspec", ref = "master")
\end{Code}

Instead of using \code{"master"}, another branch (e.\,g., \code{develop} to pull
the most recent updates) or a tag that is pointing to one of the releases
(e.\,g., \code{v1.0-0-rc1}, to install the $1$st release candidate to version
1.0-0) can be used as \code{ref}.

Note that the code in the \code{develop} branch is merged into the \code{master} branch only for a release (candidate) and after being thoroughly tested, so if you're installing from the \code{develop} branch you will potentially be using code that has not been fully tested.

Use the option \code{build_vignettes = FALSE} if you don't have \LaTeX\
installed on your system.

\subsection[R code intended for the user and its
documentation]{\proglang{R} code intended for the user and its documentation}

The classes of the \pkg{quantspec} package, their methods, slots, dependencies
and inheritance properties are implemented as conceptually designed [cf.
Section~\ref{sec:2}]. Recall that the design was presented in form of class
diagrams, on display in Figures~\ref{fig:classdiagram1}
and~\ref{fig:classdiagram2}, and that no specific programming language was
assumed. All classes that are intended for the end-user possess a constructor
method with the same name as the class itself but beginning with a lower case
letter. The classes intended for the end-user and their constructors are listed
in Table~\ref{tab:Constructors}.

\begin{table}[t]
  \centering
    \begin{tabular}{lll}
    \toprule
    Constructor & Type of object & Quantities computed \\
    \midrule
    \texttt{clippedFT}           & \texttt{ClippedFT}         & $d_n^q(\omega)$, $d_{n,R}^{\tau}(\omega)$      \\
    \texttt{qRegEstimator}       & \texttt{QRegEstimator}     & $b_n^{\tau}(\omega)$, $b_{n,R}^{\tau}(\omega)$ \\
    \texttt{quantilePG}          & \texttt{QuantilePG}        & $\hat
    L_n^{\tau_1, \tau_2}(\omega)$, $\hat L_{n,R}^{\tau_1, \tau_2}(\omega)$, \\   && $I_n^{q_1, q_2}(\omega)$, $I_{n,R}^{\tau_1, \tau_2}(\omega)$ \\
    \texttt{smoothedPG}          & \texttt{SmoothedPG}        &
    $\hat {\mathfrak{f}}_n(\tau_1, \tau_2; \omega)$, $\hat {\mathfrak{f}}_{n,R}(\tau_1,
    \tau_2; \omega)$, \\
    && $\hat G_n (q_1, q_2; \omega)$, $\hat G_{n, R} (\tau_1,
    \tau_2; \omega)$\\
    \texttt{quantileSD}          & \texttt{QuantileSD}        &
    $\mathfrak{f}_{q_1, q_2}(\omega)$, $\mathfrak{f}_{q_{\tau_1},
    q_{\tau_2}}(\omega)$ \\
    \texttt{integrQuantileSD}    & \texttt{IntegrQuantileSD}  & $\mathfrak{F}_{q_1, q_2}(\omega)$, $\mathfrak{F}_{q_{\tau_1},
    q_{\tau_2}}(\omega)$\\
    \texttt{kernelWeight}        & \texttt{KernelWeight}      & $W_n(u) = b_n^{-1} \sum_{j} W(b_n^{-1}(u + 2\pi j))$\\
    \texttt{spectrDistrWeight}   & \texttt{spectrDistrWeight} & $W_n(u) = I\{ u \leq 0 \}$ \\
    \bottomrule
    \end{tabular}%
  \caption{Constructors of the \pkg{quantspec} package that are intended for the
  end-user.} \label{tab:Constructors}%
\end{table}%

For a more detailed description of constructors and classes, documentation
within the online help system of \proglang{R} is available. After loading the
package, which is done by calling

\begin{Schunk}
\begin{Sinput}
> library("quantspec")
\end{Sinput}
\end{Schunk}

the help file of the package, which provides an overview on the design, can be
called by executing

\begin{Schunk}
\begin{Sinput}
> help("quantspec")
\end{Sinput}
\end{Schunk}

on the \proglang{R} command line.
Note that an index of all available
functions can be accessed at the very bottom of the page. If for example more
information on the constructor of \texttt{QRegEstimator} and on
the class itself is desired, then

\begin{Schunk}
\begin{Sinput}
> help("qRegEstimator")
> help("QRegEstimator")
\end{Sinput}
\end{Schunk}

should be invoked.Using this class to determine the frequency
representation $b_{n,R}^{\tau}(\omega)$, for $\tau \in \{0.25, 0.5, 0.75\}$
would look as follows. In a toy example, where eight independent random
variables $X_0, \ldots, X_7 \sim N(0,1)$ are generated and used to compute
$b_{n,R}^{\tau}(\omega)$, call

\begin{Schunk}
\begin{Sinput}
> Y  <- rnorm(8)
> bn <- qRegEstimator(Y, levels = c(0.25, 0.5, 0.75))
\end{Sinput}
\end{Schunk}

By default the computation is done for all Fourier frequencies $\omega
= 2\pi j / n \in [0,\pi]$, $n=8$, $j = 0, \ldots, \lfloor n/2 \rfloor$. The
computed information can then be viewed by typing the name of the
variable (i.\,e., \code{bn}) to the \proglang{R} console:
\begin{Schunk}
\begin{Sinput}
> bn
\end{Sinput}
\begin{Soutput}
QRegEstimator (J=5, K=3, B+1=1)
Frequencies:  0 0.7854 1.5708 2.3562 3.1416 
Levels     :  0.25 0.5 0.75 

Values:
           tau=0.25      tau=0.5      tau=0.75
0      2.000+0.000i  4.000+0.00i  6.000+0.000i
0.785 -0.354+0.354i -0.811-0.25i -0.207+1.207i
1.571  2.750+0.250i  0.250+0.75i  0.750-0.250i
2.356  0.354-0.439i  1.414+0.00i  1.207+0.207i
3.142  1.000+0.000i  1.000+0.00i  0.500+0.000i
\end{Soutput}
\end{Schunk}

Methods other than the constructor are implemented as generic functions.
To invoke the method \code{f} of an object \code{obj} the call therefore is
\code{f(obj)}.
In particular all attributes mentioned in the class diagram can be accessed via
getter methods. There are no setter methods, because all attributes are
completely handled by internal functions. For an example, to retrieve the
attributes \code{frequencies} and \code{parallel} of the object \code{bn},
execute the following lines on the \proglang{R} shell

\begin{Schunk}
\begin{Sinput}
> getFrequencies(bn)
\end{Sinput}
\begin{Soutput}
[1] 0.0000000 0.7853982 1.5707963 2.3561945 3.1415927
\end{Soutput}
\begin{Sinput}
> getParallel(bn)
\end{Sinput}
\begin{Soutput}
[1] FALSE
\end{Soutput}
\end{Schunk}

\newpage
To invoke a method \code{f} with parameters \code{p1, ..., pk} of an
object \code{obj} the call is \code{f(obj, p1, ..., pk)}. An
example is to invoke the accessor function \code{getValues}, which is equipped
with parameters to get the values associated with certain \code{frequencies} or
\code{levels}. An exemplary call looks like this:

\begin{Schunk}
\begin{Sinput}
> getValues(bn, levels = c(0.25, 0.5))
\end{Sinput}
\begin{Soutput}
, , 1

                      [,1]             [,2]
[1,]  2.0000000+0.0000000i  4.0000000+0.00i
[2,] -0.3535534+0.3535534i -0.8106602-0.25i
[3,]  2.7500000+0.2500000i  0.2500000+0.75i
[4,]  0.3535534-0.4393398i  1.4142136+0.00i
[5,]  1.0000000+0.0000000i  1.0000000+0.00i
[6,]  0.3535534+0.4393398i  1.4142136-0.00i
[7,]  2.7500000-0.2500000i  0.2500000-0.75i
[8,] -0.3535534-0.3535534i -0.8106602+0.25i
\end{Soutput}
\end{Schunk}

Note that the result is returned as an array of dimension \code{c(J, K, B + 1)},
where in the present case \code{B = 0} bootstrap replications were performed.
For a detailed description on how to use the function \code{getValues} in
the above mentioned case, access the online help via the command

\begin{Schunk}
\begin{Sinput}
> help("getValues-FreqRep")
\end{Sinput}
\end{Schunk}

Note the format \code{method_name-class_name} to access the help page of a
method and that the attribute \code{values} is part of the abstract class
\code{FreqRep} [cf. Figure~\ref{fig:classdiagram2}].

A graphical representation of the data can easily be created by applying the
\code{plot} command. For example, to compute and plot the frequency
representations $d_{32,R}^{\tau}(\omega)$, from 32 simulated, standard
normally distributed random variables execute the following lines on the
\proglang{R} shell:

\begin{Schunk}
\begin{Sinput}
> dn <- clippedFT(rnorm(32), levels = seq(0.05, 0.95, 0.05))
> plot(dn, frequencies = 2 * pi * (0:64) / 32, levels = c(0.25, 0.5))
\end{Sinput}
\end{Schunk}

\begin{figure}[t]
\begin{center}
\includegraphics{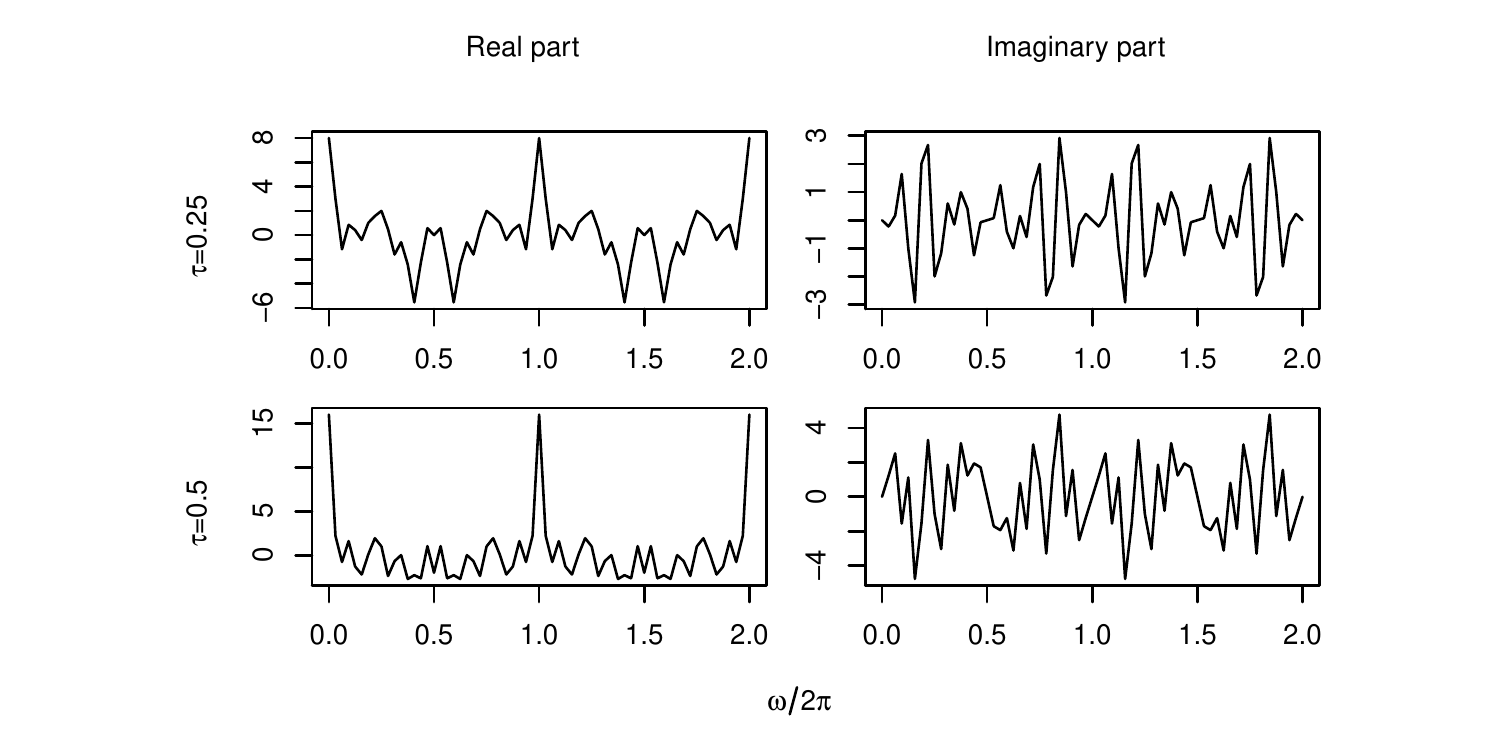}
\end{center}
  \vspace*{-0.7cm}\caption{Plot of the \code{FrequencyRepresentation} object
  \code{bn}.}
  \label{fig:plotdn}
\end{figure}

The above script will yield the diagrams that are on display in
Figure~\ref{fig:plotdn}. Note that the $d_{32,R}^{\tau}(\omega)$ were determined
for $\tau \in \{0.05, 0.1, \ldots, 0.9, 0.95\}$ and, by the default setting, for
all Fourier frequencies from $[0,\pi]$. The plot, however, was parameterized to
show only $\tau \in \{0.25, 0.5\}$, but all Fourier frequencies from $[0,4\pi]$;
by default all available \code{levels} and \code{frequencies} would be used. In
this example two of the 19 frequencies were selected to yield a plot of a
size that is apropriate to fit onto the page. Further more, the plot was
parameterized to show $d_{n,R}(\omega)$ for all Fourier frequencies from
$[0,4\pi]$ to illustrate characteristic redundancies in the frequency
representation objects, and to point out that the default values are always
sufficient. The two relations

\[d_{n,R}^{\tau}(\omega) = \overline{d_{n,R}^{\tau}(2\pi - \omega)}, \quad d_{n,R}^{\tau}(\omega) = d_{n,R}^{\tau}(\omega + 2\pi j),\]

hold for any $\omega \in \IR$ and $j \in \IZ$, $d_{n,R}^{\tau}(\omega)$.
Therefore, without additional calculations, the plot of $d_{n,R}^{\tau}(\omega)$
can be determined for any $\omega \in 2\pi j / n$, $j \in \IZ$, as long as
$d_{n,R}^{\tau}(\omega)$ is known for $\omega \in 2\pi j /n$, $j = 0,\ldots,
\lfloor n/2 \rfloor$, which is what is determined by the default setting.
Note that all of this happens transparently for the user, as the method
\code{getValues} takes care of it. Another fact that one can presume by inspecting Figure~\ref{fig:plotdn} is that $d_{n,R}^{\tau}(\omega)$ appears to be uncorrelated and centered (for $\omega \neq 0 \mod 2\pi$).

\subsection{Additional elements of the package}

The \pkg{quantspec} package includes three demos that can be accessed
via

\begin{Schunk}
\begin{Sinput}
> demo("sp500")
> demo("wheatprices")
> demo("qar-simulation")
\end{Sinput}
\end{Schunk}

Several examples explaining how to use the
various functions of the package can be found in the online help files or the
folder \code{examples} in the directory where the package is installed. The package
comes with two data sets \code{sp500} and \code{wheatprices} that are used in
the demos and in the examples.
A package vignette amends the online help files. It contains the text of this
paper.
Unit tests covering all main functions were implemented using the \pkg{testthat}
framework \citep{testthat}.

\newpage

\section{Two worked examples}
\label{sec:4}
\subsection[Analysis of the S&P 500 stock index, 2007-2010]{Analysis of the
S\&P~500 stock index, 2007--2010}
\label{sec:SP500}

In this section the use of the \pkg{quantspec} package from the perspective of a
data analysts is explained. To this end an analysis of the returns of the
S\&P~500 stock index is performed. Note that a similar analysis and the data set
used are available in the package. Calling \code{demo("sp500")} will start
the computations and by \code{sp500} the data set can be referenced to do
additional analysis.

For the example the years 2007 through to 2010 were selected to have a time
series that, at least to some degree, can be considered stationary. Aside
from this more technical consideration, employing the new statistical toolbox
will reveal interesting features in the returns collected in the financial
crisis that completely escape the analysis with the traditional tools blindly applied.

For a start, use the following \proglang{R} script to plot the data, the
autocovariances of the returns and the autocovariances of the squared returns.

\begin{Schunk}
\begin{Sinput}
> library("zoo")
> plot(sp500,            xlab = "time t", ylab = "", main = "")
> acf(coredata(sp500),   xlab = "lag k",  ylab = "", main = "")
> acf(coredata(sp500)^2, xlab = "lag k",  ylab = "", main = "")
\end{Sinput}
\end{Schunk}

\setkeys{Gin}{width=\textwidth}
\begin{figure}[t]
\begin{center}
\includegraphics{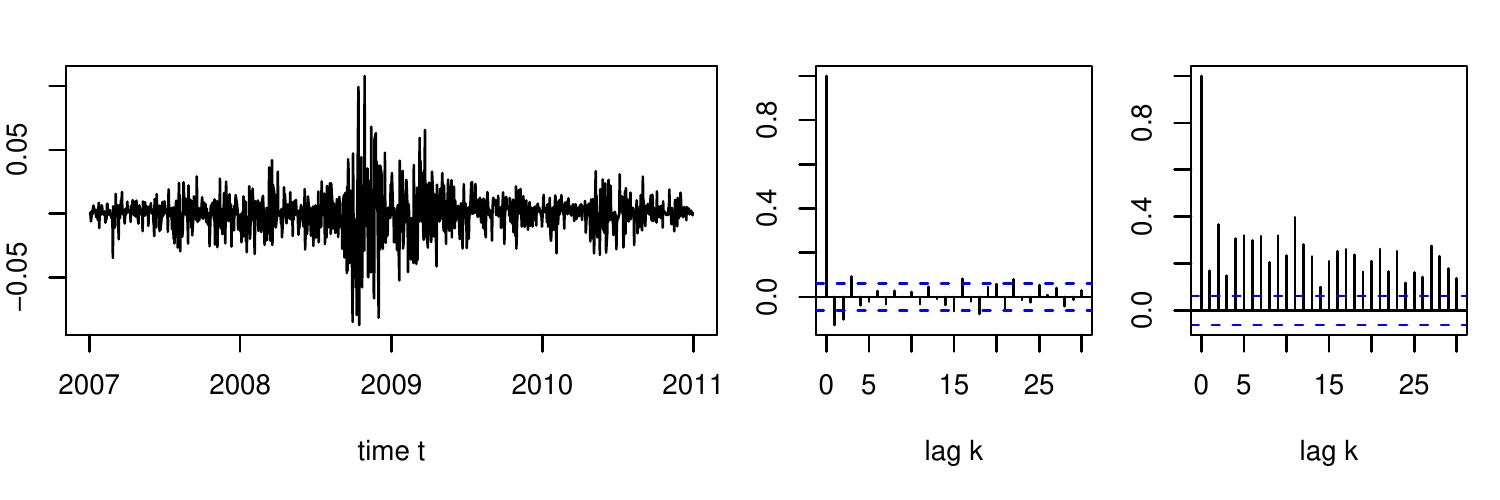}
\end{center}
  \vspace*{-0.7cm}\caption{Returns $(Y_t)$ of the S\&P~500 returns example
  data (left), autocovariances ${\rm Cov}(Y_{t+k}, Y_t)$ of the returns
  (middle), and autocovariances ${\rm Cov}(Y_{t+k}^2, Y_t^2)$ of the squared returns
  (right).}
  \label{fig:plotSP}
\end{figure}

The three plots are
displayed in Figure~\ref{fig:plotSP}. Inspecting them, it is important to
observe that the returns themselves appear to be almost uncorrelated. Therefore, not much
insight into the serial dependency structure of the data can be expected from
traditional spectral analysis. The squared returns on the other hand are
significantly correlated. This observation, typically taken as an argument to
fit an ARCH or GARCH model, clearly proves that serial dependency exists. In
what follows the copula spectral density will be estimated from the data,
using quantile periodograms and smoothing them. It will be seen that using the
\pkg{quantspec} package this can be done with only a few lines of code
necessary.

First, take a look at the CR periodogram $I_{n,R}^{\tau_1, \tau_2}(\omega)$. In
the \pkg{quantspec} package it is represented as a \code{QuantilePG} object and
can be computed calling the constructor function \code{quantilePG} with the
parameter \code{type = "clipped"}. To do the calculation for $\tau_1, \tau_2 \in
\{0.05, 0.5, 0.95\}$, all Fourier frequencies $\omega$ and with 250 bootstrap
replications determined from a moving blocks bootstrap with block length $\ell =
32$, it suffices to execute the first command of the following script:

\begin{Schunk}
\begin{Sinput}
> CR <- quantilePG(sp500, levels.1 = c(0.05, 0.5, 0.95),
+     type = "clipped", type.boot = "mbb", B = 250, l = 32)
> freq <- getFrequencies(CR)
> plot(CR, levels = c(0.05, 0.5, 0.95),
+     frequencies = freq[freq > 0 & freq <= pi],
+     ylab = expression({I[list(n, R)]^{list(tau[1], tau[2])}}(omega)))
\end{Sinput}
\end{Schunk}

\begin{figure}[p]
\begin{center}
\vspace*{-2.5cm}
\includegraphics{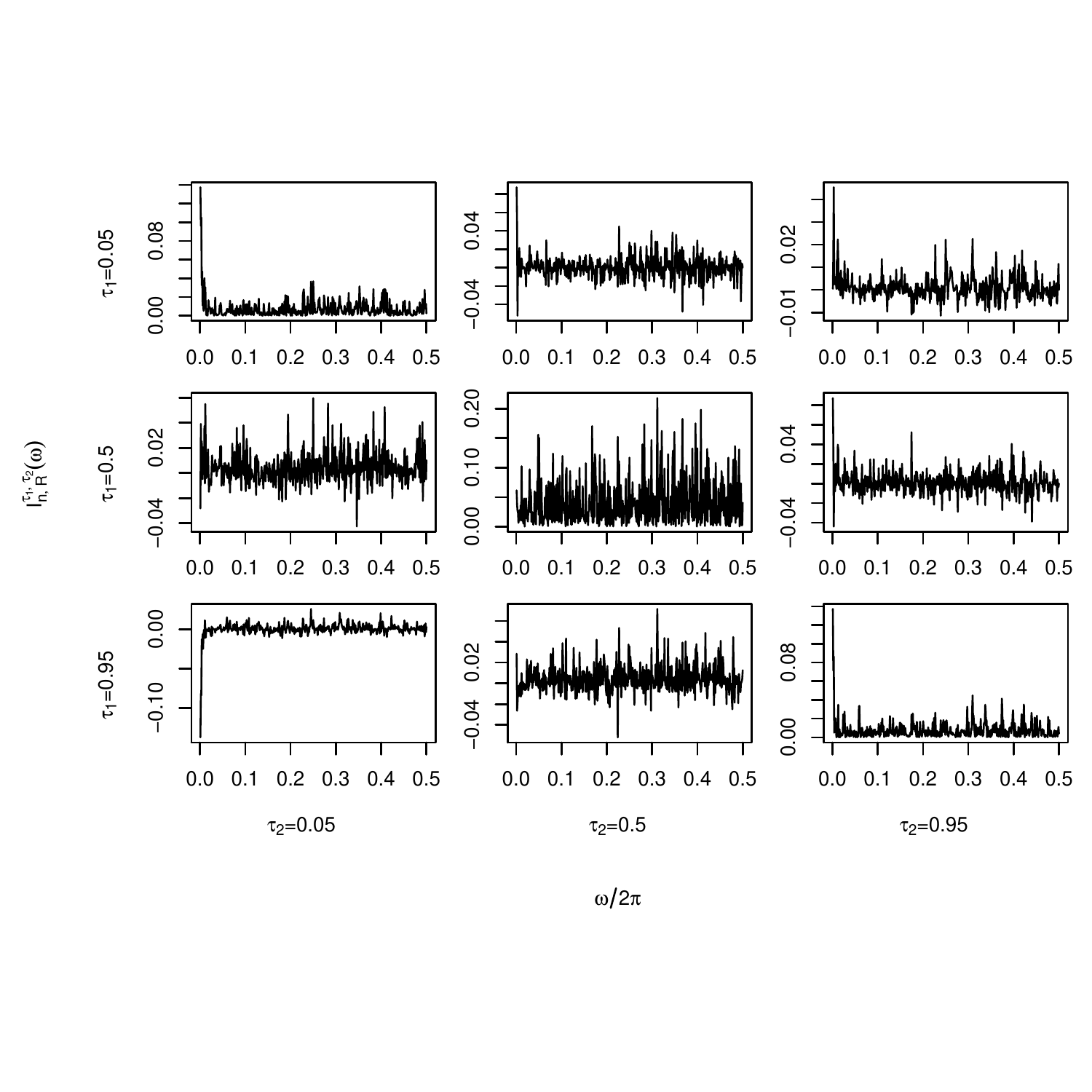}
\end{center}
  \vspace*{-2.5cm}\caption{Plot of the \code{QuantilePG} object
  \code{CR}, computed from the \code{sp500} time series;~ \protect\\
  \hspace*{1.7cm}$\omega \in (0,\pi]$.}
  \label{fig:SPquantilePG}
\begin{center}
\vspace*{-1.5cm}
\includegraphics{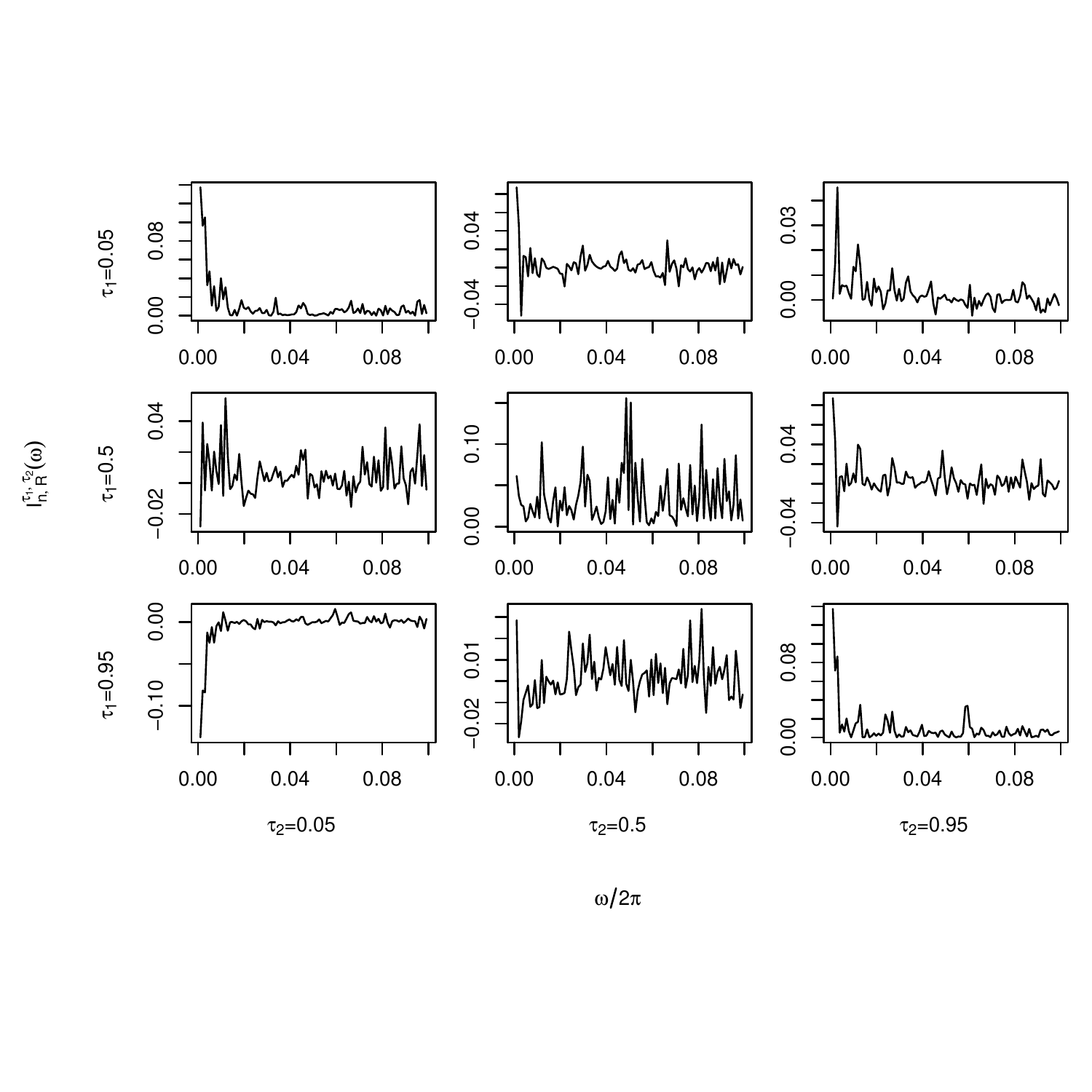}
\end{center}
  \vspace*{-2.5cm}\caption{Plot of the \code{QuantilePG} object
  \code{CR}, computed from the \code{sp500} time series;~ \protect\\
  \hspace*{1.7cm}$\omega \in (0,\pi/5]$.}
  \label{fig:SPquantilePG2}
\end{figure}

Using the second command it is possible to learn for which frequencies the
values of the CR periodogram are available. As pointed out in the previous
paragraph it was computed for all Fourier frequencies from the interval
$[0,2\pi)$, which is the default setting for \code{quantilePG} and
\code{smoothedPG}. With the third command the graphical representation of the
CR periodogram, which can be seen in Figure~\ref{fig:SPquantilePG}, is plotted.
The plot seen here is a typical plot of any \code{QSpecQuantity}: in a
configuration with $K$ levels the plot has the form of a $K \times K$ matrix,
where the subplots on and below the diagonal display the real part of the CR
periodogram $I_{n,R}^{\tau_1, \tau_2}(\cdot)$, with the levels $\tau_1$ and
$\tau_2$ denoted on the left and bottom margins of the plot. Above the diagonal
the imaginary parts are shown.

To observe the larger values in the neighborhood of $\omega = 0$ and in the
extreme quantile levels more closely a plot showing the CR periodogram
only for frequencies $\omega \in [0,\pi/5]$ can be generated using the following script:

\begin{Schunk}
\begin{Sinput}
> plot(CR, levels = c(0.05, 0.5, 0.95),
+     frequencies = freq[freq > 0 & freq <= pi/5],
+     ylab = expression({I[list(n, R)]^{list(tau[1], tau[2])}}(omega)))
\end{Sinput}
\end{Schunk}

The plot is shown in Figure~\ref{fig:SPquantilePG2}.

In the next step the computed quantile periodogram \code{CR} can be used as the
basis to determine a smoothed CR periodogram \code{sCR}. In the form of a
\code{SmoothedPG} object it can be generated by the constructor
\code{smoothedPG} of that class. Besides the \code{QuantilePG} object \code{CR},
a \code{KernelWeight} object is required, which is easily generated using
the constructor \code{kernelWeight}. As parameters the constructor
\code{kernelWeight} requires a kernel \code{W} and a bandwidth \code{bw}. The
\pkg{quantspec} package comes with several kernels already implemented. The
Epanechnikov kernel for example can be refered to by the name \code{W1}. For a
complete list of the available kernels call

\begin{Schunk}
\begin{Sinput}
> help("kernels")
\end{Sinput}
\end{Schunk}

\pagebreak
To compute the smoothed CR
periodogram from \code{CR} using the Epanechnikov kernel and bandwidth \code{bw
= 0.07} the first of the following two commands need to be executed.

\begin{Schunk}
\begin{Sinput}
> sPG <- smoothedPG(CR, weight = kernelWeight(W = W1, bw = 0.07))
> plot(sPG, levels = c(0.05, 0.5, 0.95), type.scaling = "individual",
+     frequencies = freq[freq > 0 & freq <= pi], ptw.CIs = 0.1,
+     ylab = expression(hat(G)[list(n, R)](list(tau[1], tau[2], omega))))
\end{Sinput}
\end{Schunk}

\begin{figure}[p]
\begin{center}
\vspace*{-2.5cm}
\includegraphics{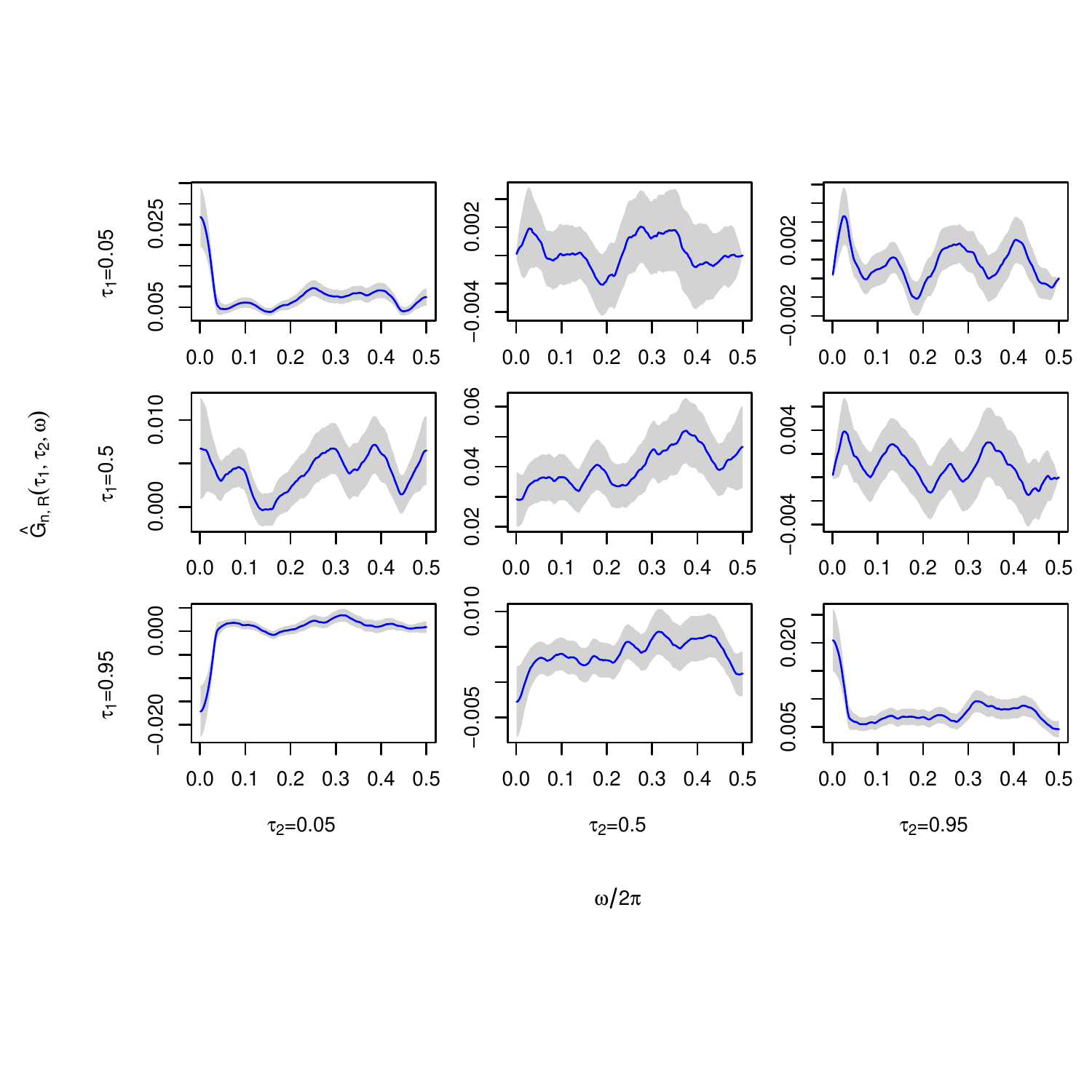}
\end{center}
  \vspace*{-2.5cm}\caption{Plot of the \code{SmoothedPG} object
  \code{sCR}, computed from the \code{sp500} time series;\protect\\
  \hspace*{1.7cm}\code{type.scaling = "individual"}, \code{ptw.CIs = 0.1}.}
  \label{fig:SPsmoothedPG}

\begin{center}
\vspace*{-1.5cm}
\includegraphics{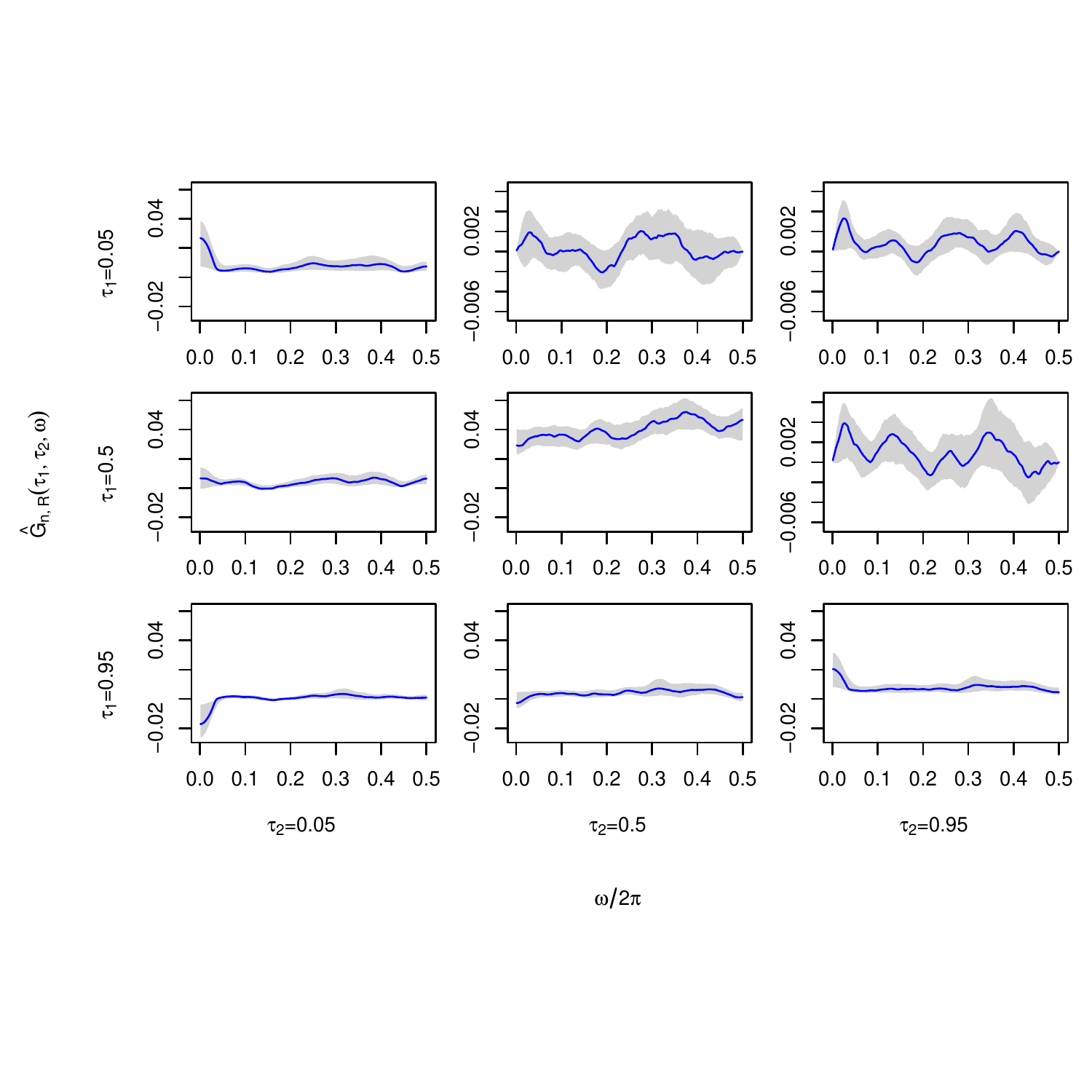}
\end{center}
  \vspace*{-2.5cm}\caption{Plot of the \code{SmoothedPG} object
  \code{sCR}, computed from the \code{sp500} time
  series;\protect\\
  \hspace*{1.7cm}\code{type.scaling = "real-imaginary"}, \code{ptw.CIs =
  0.1},\protect\\
  \hspace*{1.7cm}\code{type.CIs = "boot.full"}.}
  \label{fig:SPsmoothedPGboot}
\end{figure}

Of course, the second line initiates plotting the smoothed CR periodogram, which
is on display in Figure~\ref{fig:SPsmoothedPG}. Note that the option
\code{type.scaling} can be set to yield a plot with certain subplots possessing
the same scale. In Figure~\ref{fig:SPsmoothedPG} pointwise confidence intervals
are shown. By default these are determined using a normal approximation to the
distribution of the estimator as is suggested by the limit theorem in
\cite{KleyEtAl2014}. An alternative is to use the quantiles of estimates
computed from the block bootstrap replicates. These pointwise confidence
intervals can be plotted using the option \code{type.CIs = "boot.full"}, as is
shown in the following script:

\begin{Schunk}
\begin{Sinput}
> plot(sPG, levels = c(0.05, 0.5, 0.95), type.scaling = "real-imaginary",
+     ptw.CIs = 0.1, type.CIs = "boot.full",
+     frequencies = freq[freq > 0 & freq <= pi],
+     ylab = expression(hat(G)[list(n, R)](list(tau[1], tau[2], omega))))
\end{Sinput}
\end{Schunk}

For illustrative purposes a different type of scaling was used for the second
plot. A complete description of the options is available in the online
help, which can be accessed by calling

\begin{Schunk}
\begin{Sinput}
> help("plot-SmoothedPG")
\end{Sinput}
\end{Schunk}

Inspecting the plots in Figures~\ref{fig:SPsmoothedPG}
and~\ref{fig:SPsmoothedPGboot} reveals that serial dependency in the events
$\{Y_t \leq q_{0.05}\}$ and $\{Y_t \leq q_{0.95}\}$ is present in the data. This
concludes the introduction of the \pkg{quantspec} package for data analysts and
we can continue with the presentation of how it can also make the work of a
probability theorist easier.

\clearpage

\setkeys{Gin}{width=0.9\textwidth}

\setkeys{Gin}{width=0.9\textwidth}
\setkeys{Gin}{width=\textwidth}

%\FloatBarrier

\clearpage
\subsection{A simulation study: Analysing a quantile autoregressive
process}

In this section using the \pkg{quantspec} package from the perspective of a
probability theorist is explained. The aim is twofold. On the one hand,
further insight into a stochastic process shall be gained. Any process for which a function to simulate
finite stretches of is available can be studied. On the other hand, the finite
sample performance of the new spectral methods are to be evaluated. Note that
the example discussed in this section and the functions to simulate QAR(1)
processes are available inside the package, by calling
\code{demo("qar-simulation")} and by referring to the function \code{ts1}, which
implements the QAR(1) process that was discussed in \cite{DetteEtAl2013} and
\cite{KleyEtAl2014}. Recall that a QAR(1) process is a sequence $(X_t)$ of
random variables that fulfills
\[X_t = \theta_1(U_t) X_{t-1} + \theta_0(U_t),\]
where $U_t$ is independent white noise with $U_t \sim \mathcal{U}[0,1]$, and
$\theta_1, \theta_0: [0,1] \rightarrow \IR$ are model parameters
\citep{Koenker2006}. The function \code{ts1} implements the model, where
$\theta_1(u) = 1.9 (u-0.5)$, $u \in [0,1]$ and $\theta_0 = \Phi^{-1}$, which was
discussed in \cite{DetteEtAl2013} and \cite{KleyEtAl2014}. A complete list of
models included in the package can be seen in the online documentation of the
package by calling

\begin{Schunk}
\begin{Sinput}
> help("ts-models")
\end{Sinput}
\end{Schunk}

\begin{figure}[p]
\begin{center}
\vspace*{-2.5cm}
\includegraphics{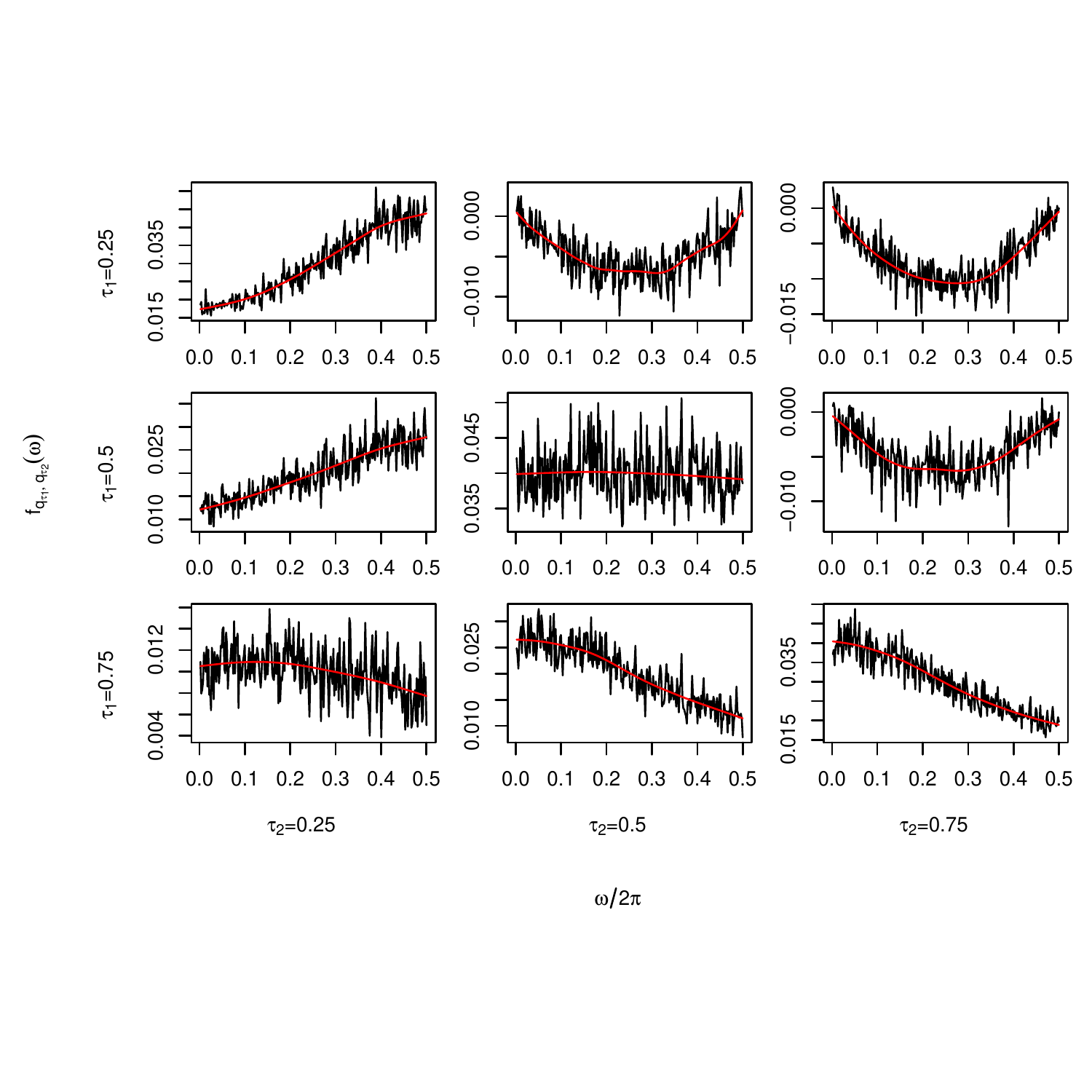}
\end{center}
  \vspace*{-2.5cm}\caption{Plot of the copula spectral density
  $\mathfrak{f}_{q_{\tau_1}, q_{\tau_2}}(\omega)$ of the QAR(1) model;\protect\\
  \hspace*{1.7cm} $\tau_1, \tau_2 \in \{0.25,0.5,0.75\}$, and $\omega \in [0,\pi]$; $N = 2^9$ and $R = 100$.}
  \label{fig:csdQAR}

\begin{center}
\vspace*{-1.5cm}
\includegraphics{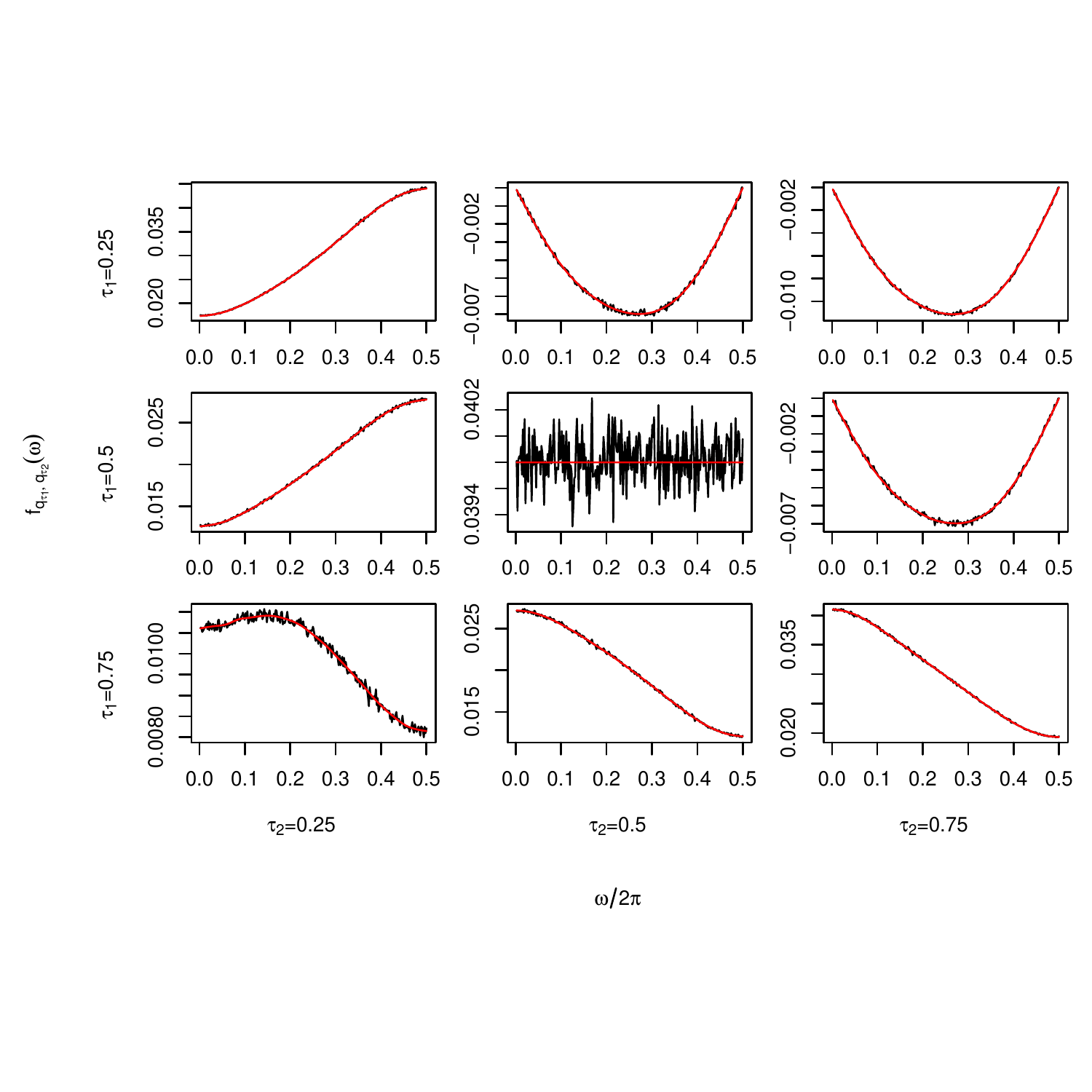}
\end{center}
  \vspace*{-2.5cm}\caption{Plot of the copula spectral density
  $\mathfrak{f}_{q_{\tau_1}, q_{\tau_2}}(\omega)$ of the QAR(1) model;\protect\\
  \hspace*{1.7cm} $\tau_1, \tau_2 \in \{0.25,0.5,0.75\}$, and $\omega \in [0,\pi]$; $N = 2^{12}$ and $R = 50000$.}
  \label{fig:csdQARhighprecPlot}
\end{figure}

The following, two-line script can be used to generate the graphical
representation of the copula spectral density that is on display in Figure~\ref{fig:csdQAR}

\begin{Schunk}
\begin{Sinput}
> csd <- quantileSD(N = 2^9, seed.init = 2581, type = "copula",
+         ts = ts1, levels.1 = c(0.25, 0.5, 0.75), R = 100, quiet = TRUE)
> plot(csd, ylab = expression(f[list(q[tau[1]], q[tau[2]])](omega)))
\end{Sinput}
\end{Schunk}

When analysing a time series model the recommended practice is to compute
the quantile spectral density once with high precision, store it to the
hard drive, and load it later whenever it is needed. The following two lines of
code can be used to do this:

\begin{Schunk}
\begin{Sinput}
> csd <- quantileSD(N=2^12, seed.init = 2581, type = "copula",
+     ts = ts1, levels.1 = c(0.25, 0.5, 0.75), R = 50000)
> save(csd, file="csd-qar1.rdata")
\end{Sinput}
\end{Schunk}

With the first configuration ($N=2^9$ and $R=100$) the computation time
was only around three seconds. To compute the second \code{csd} object (with
$N=2^{12}$ and $R=50000$) the same machine needed roughly 2.5 hours. Storing the
object in a file takes about 1MB of hard disk space. Not only
\code{values} and \code{stdError}s are stored within the \code{QuantileSD}
object; also the final state of the pseudo random number generator is stored and
the method \code{increasePrecision} can be used to add more simulation runs at
any time to yield a better approximation to the true quantile spectrum. More
information on this method can be found in the online help, which is accessible
via

\begin{Schunk}
\begin{Sinput}
> help("increasePrecision-QuantileSD")
\end{Sinput}
\end{Schunk}

Once the computation is finished the diagram on display in
Figure~\ref{fig:csdQARhighprecPlot} can be created using the following two lines
of code:

\begin{Schunk}
\begin{Sinput}
> load("csd-qar1.rdata")
> plot(csd, frequencies = 2 * pi * (1:2^8) / 2^9,
+     ylab = expression(f[list(q[tau[1]], q[tau[2]])](omega)))
\end{Sinput}
\end{Schunk}

The parameter \code{frequencies} was used when plotting the copula spectral
density to create a plot that can be compared to the one in Figure~\ref{fig:csdQAR}.
Note that by default the plot would have been created using all available
frequencies which yields a grid of 8 times as many points on the x-axis
($N=2^{12}$ vs. $N=2^9$). Now, to get a first idea of how well the estimator
performs, plot the smoothed CR periodogram computed from one simulated QAR(1)
time series of length 512:

\begin{Schunk}
\begin{Sinput}
> sCR <- smoothedPG(ts1(512), levels.1 = c(0.25, 0.5, 0.75),
+     weight = kernelWeight(W = W1, bw = 0.1))
> plot(sCR, qsd = csd,
+     ylab = bquote(paste(hat(G)[list(n, R)](list(tau[1], tau[2], omega)),
+     " and ", f[list(q[tau[1]], q[tau[2]])](omega))))
\end{Sinput}
\end{Schunk}

\begin{figure}[h!]
\begin{center}
\vspace*{-2.5cm}
\includegraphics{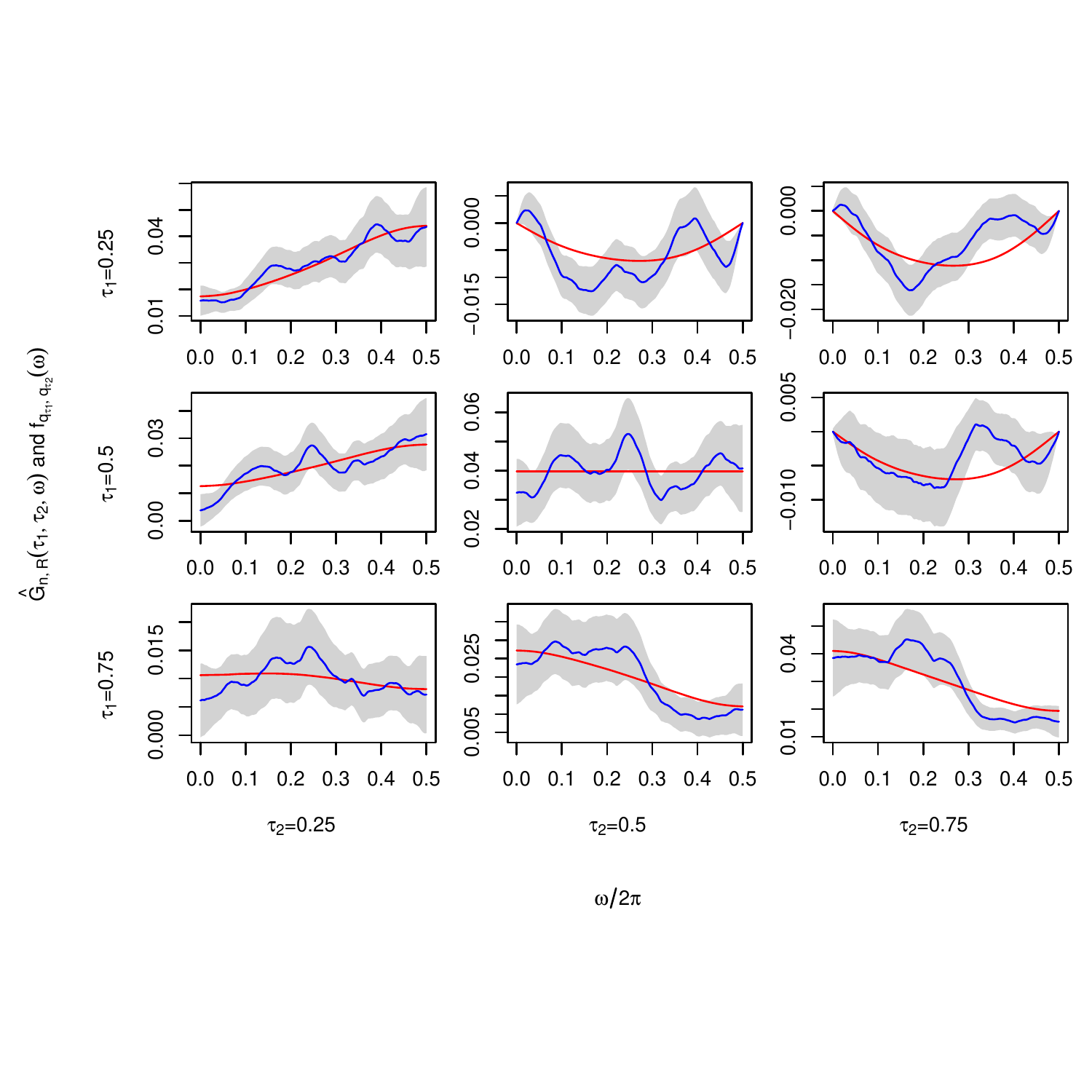}
\end{center}
  \vspace*{-2.5cm}\caption{Plot of a smoothed CR periodogram,\protect\\
  \hspace*{1.7cm} computed from one realization of a QAR(1) time
  series;\protect\\
  \hspace*{1.7cm} $n=512$, Epanechnikov kernel with $b_n = 0.1$.}
  \label{fig:oneQAR}
\end{figure}

The generated plot is on display in Figure~\ref{fig:oneQAR}. It is worth
pointing out that in this example ($N=512$) the estimator performs already quite
well.
Note that a different version of the constructor \code{smoothedPG} was used here
than in Section~\ref{sec:SP500}. When computing a smoothed quantile periodogram
straight from a time series, the syntax is the same as for \code{quantilePG},
but with the additional parameter \code{weight}.

Finally, for the simulation study, a number of \code{R = 5000} independent
QAR(1) time series are generated. Before the actual simulations, some
variables that determine what is to be simulated are defined:

\begin{Schunk}
\begin{Sinput}
> set.seed(2581)
> ts <- ts1
> N <- 128
> R <- 5000
> freq <- 2  * pi * (1:16) / 32
> levels <- c(0.25, 0.5, 0.75)
> J <- length(freq)
> K <- length(levels)
> sims  <- array(0, dim=c(4, R, J, K, K))
> weight <- kernelWeight(W = W1, bw = 0.3)
\end{Sinput}
\end{Schunk}

Setting the seed in the very beginning allows for reproducible results.
Recall that \code{ts1} is a function to simulate from the QAR(1) model to be
studied. \code{N} is the length of the time series and also the number of
Fourier frequencies for which the quantile periodograms will be computed. By the
parameter \code{freq} a subset of these Fourier frequencies is specified to be
stored; a subset is used to save storage space. The estimates at these
frequencies \code{freq} and at the specified \code{levels} are then stored to
the array \code{sims}. In this example, the smoothed periodograms are computed
using the Epanechnikov kernel and the (rather large) bandwidth of $b_n = 0.3$.

\clearpage

\setkeys{Gin}{width=0.9\textwidth}
\setkeys{Gin}{width=\textwidth}

\clearpage

\setkeys{Gin}{width=0.9\textwidth}

%\FloatBarrier

For the actual simulation the following \code{for} loop can be used:

\begin{Schunk}
\begin{Sinput}
> for (i in 1:R) {
+   Y <- ts(N)
+ 
+   CR  <- quantilePG(Y, levels.1=levels, type = "clipped")
+   LP  <- quantilePG(Y, levels.1=levels, type = "qr")
+   sCR <- smoothedPG(CR, weight = weight)
+   sLP <- smoothedPG(LP, weight = weight)
+ 
+   sims[1, i, , , ] <- getValues(CR,  frequencies=freq)[, , , 1]
+   sims[2, i, , , ] <- getValues(LP,  frequencies=freq)[, , , 1]
+   sims[3, i, , , ] <- getValues(sCR, frequencies=freq)[, , , 1]
+   sims[4, i, , , ] <- getValues(sLP, frequencies=freq)[, , , 1]
+ }
\end{Sinput}
\end{Schunk}

Note that the flexible accessor method \code{getValues} is used to access the
relevant subset of values for the frequencies specified (i.\,e., \code{freq}).
Once the array \code{sims} is available, many interesting properties of
the estimator can be analyzed. Examples include the bias,
variance, etc. Here, using the function \code{getValues} again, the true copula
spectral density is copied to an array \code{trueV}. Using the
arrays \code{sims} and \code{trueV} the root integrated mean squared errors are
computed as follows:

\begin{Schunk}
\begin{Sinput}
> trueV <- getValues(csd, frequencies = freq)
> SqDev <- array(apply(sims, c(1, 2),
+         function(x) {abs(x - trueV)^2}), dim=c(J, K, K, 4, R))
> rimse <- sqrt(apply(SqDev, c(2, 3, 4), mean))
\end{Sinput}
\end{Schunk}

\begin{Schunk}
\begin{Sinput}
> rimse
\end{Sinput}
\begin{Soutput}
, , 1

           [,1]       [,2]       [,3]
[1,] 0.03292733 0.03543752 0.02879294
[2,] 0.03543752 0.04113916 0.03427386
[3,] 0.02879294 0.03427386 0.03014753

, , 2

           [,1]       [,2]       [,3]
[1,] 0.02688778 0.03275136 0.02691644
[2,] 0.03275136 0.03447488 0.03191837
[3,] 0.02691644 0.03191837 0.02526850

, , 3

            [,1]        [,2]        [,3]
[1,] 0.004338658 0.005889695 0.004472232
[2,] 0.005889695 0.006794010 0.005428915
[3,] 0.004472232 0.005428915 0.004917286

, , 4

            [,1]        [,2]        [,3]
[1,] 0.005133067 0.005699958 0.004575845
[2,] 0.005699958 0.006179409 0.005386339
[3,] 0.004575845 0.005386339 0.004797662
\end{Soutput}
\end{Schunk}

These numbers could now be inspected to observe, for example, that the smoothed
quantile periodogram possess smaller root integrated mean squared errors than the
quantile periodograms (without smoothing). Further discussion of the numbers is
omitted, because the crux of this chapter was to explain how to perform the
simulation study, not to actually do it.

\section{Roadmap to future developments and concluding remarks}

As the new methodology evolves additional features will be added to the
\pkg{quantspec} package. For each new feature an entry to the issue tracker
available on the GitHub will be made. Then the new feature will be implemented
on a topic branch of the repository. For example, a procedure for data-driven
selection of the bandwidth is currently being developed.

Other, more complex extensions to the software include the implementation of
functions to perform quantile spectral analysis for locally stationary
processes, the computation and smoothing of higher order quantile periodograms for
the estimation of quantile polyspectra. Procedures for graphical representations
of these objects, possibly animated ones, will amend these planned parts of the
package. An overview on the planned extensions will be made available in the
issue tracker on GitHub.

Summing up, it can be said that the \pkg{quantspec} package provides a
comprehensive and conclusive toolbox to perform quantile-based spectral
analysis. Due to the great interest in and active development of the
statistical procedures that are quantile-based spectral analysis it was
deliberately designed in an object-oriented and extensible fashion. Thus it is
well prepared for the many extensions that are sure to come in the near future.
The source code is open and extensive documentation of the system freely
available. Comments on and contribution to the project is, of course, very much
welcome.

\section*{Acknowledgments}
This work has been supported by the Sonderforschungsbereich ``Statistical
modeling of nonlinear dynamic processes'' (SFB 823) of the Deutsche
Forschungsgemeinschaft and by a PhD Grant of the Ruhr-Universit\"{a}t Bochum
and by the Ruhr- Universit\"{a}t Research School funded by Germany's Excellence Initiative [DFG
GSC 98/1].

%\FloatBarrier

\bibliography{paper-quantspec}

\begin{thebibliography}{25}
\newcommand{\enquote}[1]{``#1''}
\providecommand{\natexlab}[1]{#1}
\providecommand{\url}[1]{\texttt{#1}}
\providecommand{\urlprefix}{URL }
\expandafter\ifx\csname urlstyle\endcsname\relax
  \providecommand{\doi}[1]{doi:\discretionary{}{}{}#1}\else
  \providecommand{\doi}{doi:\discretionary{}{}{}\begingroup
  \urlstyle{rm}\Url}\fi
\providecommand{\eprint}[2][]{\url{#2}}

\bibitem[{Brillinger(1975)}]{Brillinger1975}
Brillinger DR (1975).
\newblock \emph{Time Series: Data Analysis and Theory}.
\newblock Holt, Rinehart and Winston, Inc.

\bibitem[{Dette \emph{et~al.}(2014+)Dette, Hallin, Kley, and
  Volgushev}]{DetteEtAl2013}
Dette H, Hallin M, Kley T, Volgushev S (2014+).
\newblock \enquote{Of Copulas, Quantiles, Ranks and Spectra: An $L_1$-Approach
  to Spectral Analysis.}
\newblock \emph{Bernoulli}, \textbf{forthcoming}.

\bibitem[{Hagemann(2011)}]{Hagemann2011}
Hagemann A (2011).
\newblock \enquote{Robust Spectral Analysis.}
\newblock \emph{arXiv preprint}.
\newblock \urlprefix\url{http://arxiv.org/abs/1111.1965}.

\bibitem[{Hill and McCloskey(2013)}]{HillMcCloskey2013}
Hill JB, McCloskey A (2013).
\newblock \enquote{Heavy Tail Robust Frequency Domain Estimation.}

\bibitem[{Hong(1999)}]{Hong1999}
Hong Y (1999).
\newblock \enquote{Hypothesis Testing in Time Series via the Empirical
  Characteristic Function: A Generalized Spectral Density Approach.}
\newblock \emph{Journal of the American Statistical Association},
  \textbf{94}(448), 1201--1220.

\bibitem[{Hong(2000)}]{Hong2000}
Hong Y (2000).
\newblock \enquote{Generalized Spectral Tests for Serial Dependence.}
\newblock \emph{Journal of the Royal Statistical Society B}, \textbf{62}(3),
  557--574.

\bibitem[{Katkovnik(1998)}]{Katkovnik1998}
Katkovnik V (1998).
\newblock \enquote{Robust {M}-Periodogram.}
\newblock \emph{IEEE Transactions on Signal Processing}, \textbf{46}(11),
  3104--3109.

\bibitem[{Kedem(1980)}]{Kedem1980}
Kedem B (1980).
\newblock \emph{Binary Time Series}.
\newblock Dekker, New York.

\bibitem[{Kleiner \emph{et~al.}(1979)Kleiner, Martin, and
  Thomson}]{KleinerEtAl1979}
Kleiner B, Martin RD, Thomson DJ (1979).
\newblock \enquote{Robust Estimation of Power Spectra.}
\newblock \emph{Journal of the Royal Statistical Society B}, \textbf{41}(3),
  313--351.

\bibitem[{Kley(2014{\natexlab{a}})}]{Kley2014}
Kley T (2014{\natexlab{a}}).
\newblock \emph{\pkg{quantspec}: Quantile-Based Spectral Analysis Functions}.
\newblock \proglang{R}~package version~\mbox{1.0-1},
  \urlprefix\url{http://CRAN.R-project.org/package=quantspec}.

\bibitem[{Kley(2014{\natexlab{b}})}]{Kley2014a}
Kley T (2014{\natexlab{b}}).
\newblock \emph{Quantile-Based Spectral Analysis: Asymptotic Theory and
  Computation}.
\newblock {PhD Thesis}, Ruhr University Bochum.
\newblock
  \urlprefix\url{http://www-brs.ub.rub.de/netahtml/HSS/Diss/KleyTobias/}.

\bibitem[{Kley \emph{et~al.}(2014)Kley, Volgushev, Dette, and
  Hallin}]{KleyEtAl2014}
Kley T, Volgushev S, Dette H, Hallin M (2014).
\newblock \enquote{Quantile Spectral Processes: Asymptotic Analysis and
  Inference.}
\newblock \emph{arXiv preprint}.
\newblock \urlprefix\url{http://arxiv.org/abs/1401.8104}.

\bibitem[{Kl{\"u}ppelberg and Mikosch(1994)}]{KluppelbergMikosch1994}
Kl{\"u}ppelberg C, Mikosch T (1994).
\newblock \enquote{Some Limit Theory for the Self-Normalised Periodogram of
  Stable Processes.}
\newblock \emph{Scandinavian Journal of Statistics}, pp. 485--491.

\bibitem[{Koenker(2005)}]{Koenker2005}
Koenker R (2005).
\newblock \emph{Quantile Regression}.
\newblock Econometric Society Monographs. Cambridge University Press.

\bibitem[{Koenker(2013)}]{quantreg}
Koenker R (2013).
\newblock \emph{\pkg{quantreg}: Quantile Regression}.
\newblock \proglang{R}~package version 5.05,
  \urlprefix\url{http://CRAN.R-project.org/package=quantreg}.

\bibitem[{Koenker and Xiao(2006)}]{Koenker2006}
Koenker R, Xiao Z (2006).
\newblock \enquote{Quantile Autoregression.}
\newblock \emph{Journal of the American Statistical Association},
  \textbf{101}(475), 980--990.

\bibitem[{Lee and Rao(2012)}]{LeeRao2012}
Lee J, Rao SS (2012).
\newblock \enquote{The Quantile Spectral Density and Comparison Based Tests for
  Nonlinear Time Series.}
\newblock \emph{arXiv preprint}.
\newblock \urlprefix\url{http://arxiv.org/abs/1112.2759}.

\bibitem[{Li(2008)}]{Li2008}
Li TH (2008).
\newblock \enquote{Laplace Periodogram for Time Series Analysis.}
\newblock \emph{Journal of the American Statistical Association},
  \textbf{103}(482), 757--768.

\bibitem[{Li(2012)}]{Li2012}
Li TH (2012).
\newblock \enquote{Quantile Periodograms.}
\newblock \emph{Journal of the American Statistical Association},
  \textbf{107}(498), 765--776.

\bibitem[{Li(2013)}]{Li2013}
Li TH (2013).
\newblock \emph{Time Series with Mixed Spectra: Theory and Methods}.
\newblock CRC Press, Boca Raton.

\bibitem[{Li(2014)}]{Li2014}
Li TH (2014).
\newblock \enquote{Quantile Periodogram and Time-Dependent Variance.}
\newblock \emph{Journal of Time Series Analysis}, \textbf{35}(4), 322--340.

\bibitem[{Mikosch(1998)}]{Mikosch1998}
Mikosch T (1998).
\newblock \enquote{Periodogram Estimates from Heavy-Tailed Data.}
\newblock In RA~Adler, R~Feldman, MS~Taqqu (eds.), \emph{A Practical Guide to
  Heavy Tails: Statistical Techniques for Analysing Heavy-Tailed
  Distributions}, pp. 241--258. Birkh{\"a}user, Boston.

\bibitem[{{\proglang{R} Core Team}(2012)}]{R2012}
{\proglang{R} Core Team} (2012).
\newblock \emph{\proglang{R}: A Language and Environment for Statistical
  Computing}.
\newblock \proglang{R}~Foundation for Statistical Computing, Vienna, Austria.
\newblock {ISBN} 3-900051-07-0, \urlprefix\url{http://www.R-project.org/}.

\bibitem[{Wickham(2011)}]{testthat}
Wickham H (2011).
\newblock \enquote{\pkg{testthat}: Get Started with Testing.}
\newblock \emph{The \proglang{R}~Journal}, \textbf{3}, 5--10.
\newblock
  \urlprefix\url{http://journal.r-project.org/archive/2011-1/RJournal_2011-1_Wickham.pdf}.

\bibitem[{Wickham and Chang(2013)}]{devtools}
Wickham H, Chang W (2013).
\newblock \emph{\pkg{devtools}: Tools to Make Developing \proglang{R} Code
  Easier}.
\newblock \proglang{R}~package version 1.4.1,
  \urlprefix\url{http://CRAN.R-project.org/package=devtools}.

\end{thebibliography}

\end{document}